\documentclass[twocolumn]{aastex631}

\newcommand{\lum}{erg s$^{\rm -1}$\,}
\newcommand{\density}{cm$^{\rm -3}$\,}
\newcommand{\vol}{cm$^{\rm 3}$\,}
\newcommand{\pressure}{dyne cm$^{\rm -2}$\,}

\newcommand{\lxf}{ L$_{\rm XF}$\,}

\newcommand{\nh}{{\rm $N_H$}\,}
\usepackage[T1]{fontenc}
\usepackage{newtxtext,newtxmath,subfigure}

\usepackage{tablefootnote}

\begin{document}

\title[]{Characterizing Superflares in HR 1099 using Temporal and Spectral Analysis of XMM-Newton Observations}
\shorttitle{X-ray Superflares from HR 1099}
 \shortauthors{Didel et al.}

\author{Shweta Didel}
\affiliation{Department of Physics, Indian Institute of Technology (BHU), \\
Varanasi-221005, India \\}

\author{Jeewan C. Pandey}
\affiliation{ Aryabhatta Research Institute of Observational Sciences (ARIES), \\
Manora Peak, Nainital-263001, India \\}


\author{A.K.~Srivastava}
\affiliation{Department of Physics, Indian Institute of Technology (BHU), \\
Varanasi-221005, India \\}

\correspondingauthor{Shweta Didel}
\email{shwetadidel@gmail.com}

\begin{abstract}
In the present paper, we analyze three energetic X-ray flares from the active RS CVn binary HR 1099 using data obtained from XMM-Newton. The flare duration ranges from 2.8 to 4.1 h, with e-folding rise and decay times in the range of 27 to 38 minutes and 1.3 to 2.4 h, respectively, indicating rapid rise and slower decay phases. The flare frequency for HR 1099 is one flare per rotation period. Time-resolved spectroscopy reveals peak flare temperatures of 39.44, 35.96, and 32.48 MK, emission measures of $7 \times 10^{53}$ to $8 \times 10^{54}$ cm$^{-3}$, global abundances of 0.250, 0.299, and 0.362 $Z_\odot$, and peak X-ray luminosities of $ 10^{31.21-32.29}$ erg s$^{-1}$. The quiescent state is modeled with a three-temperature plasma maintained at 3.02, 6.96, and 12.53 MK. Elemental abundances during quiescent and flaring states exhibit the inverse-FIP effect. We have conducted a comparative analysis of coronal abundances with previous studies and found evidence supporting the i-FIP effect. The derived flare semi-loop lengths of 6 to 8.9 $\times 10^{10}$ cm were found to be comparable to the other flares detected on HR 1099; however, they are significantly larger than typical solar flare loops. The estimated flare energies, ranging from $10^{35.83-37.03}$ erg, classify these flares as super-flares. The magnetic field strengths of the loops are found to be in the range of 350 to 450 G. We diagnose the physical conditions of the flaring corona in HR 1099 through the observations of superflares and provide inference on the plasma processes.
\end{abstract}


\section{Introduction}
\label{sec:intro}
Solar and stellar flares are the transient events resulting from the reconnection of magnetic loops in conjunction with the emission of radiation and particle beams, along with chromospheric evaporation, rapid mass flows, and plasma heating.
From an observational perspective, flares manifest across the electromagnetic spectrum. They are characterized by a rapid rise in radiation flux, peaking at different times in various wavelengths, subsequently followed by a slow decay phase. The flares observed in the RS CVn type binary exhibit numerous similarities to the solar and other stellar flares \citep{2008MNRAS.387.1627P,2012MNRAS.419.1219P}. Nonetheless, significant differences exist in their activity indicators, including variations in the magnitude of released energy from the flaring events in the different classes of active stars. The tidal interaction between components of RS CVn binary systems can synchronize their rotational and orbital periods. Moreover, the expanded convection zone in the evolved giant or subgiant component results in much stronger magnetic activity in RS CVn binaries compared to the other late-type stars as well as the Sun. Therefore, studying flares in these cool giants and subgiants is particularly crucial, as it offers new insight into the dynamic behavior of the corona of RS CVns, which may differ significantly from that of dwarf stars.

HR 1099 (= V711 Tauri) is the RS CVn type binary \citep{1976AJ.....81..771B} located in the Taurus constellation at a distance of $29.43\pm0.03$ pc \citep{2020yCat.1350....0G}. Comprising a pair of solar-type stars, HR 1099 is composed of a K1IV star as primary and a smaller, rapidly rotating G5V secondary. This system exhibits a short orbital period of approximately 2.82 days \citep[][]{1976AJ.....81..771B,1983ApJ...268..274F}, resulting in close proximity between these two stars. One of the key features that sets HR 1099 apart is its exceptional magnetic activity. Both stars in the system generate strong magnetic fields, which in turn give rise to various physical phenomena, including surface inhomogeneity, stellar flares, X-ray emissions, and other forms of high-energy radiations, including gamma and UV radiations \citep[e.g.][]{2003A&A...397..285G,2004ApJS..153..317O}. This makes HR 1099 an ideal target for studying the complex magnetic activities resulting from X-rays and other wavelengths.

HR 1099 has been extensively observed in the past from X-ray to radio wavelengths \citep[e.g.][]{2004ApJS..153..317O,2003A&A...397..285G,2001ApJ...549..554A,2000A&A...354..537S}.  
Many authors have done extensive analysis to identify magnetic activity cycles in HR 1099 using long-term photometry, a method known for its ability to detect changes in spot coverage. These investigations have yielded varying periods ranging from 14 -- 20 yrs \citep{2010A&A...521A..36M, 2007ApJ...659L.157B,2006A&A...455..595L}. \cite{2018A&A...616A.161P} explored the long-term X-ray activity cycle and found no statistically significant periodic variations in the X-ray dataset. HR 1099 is prone to X-ray flaring events \citep[e.g.][]{2017ATel10753....1K,2014GCN.15679....1B,2012MNRAS.419.1219P,2007A&A...464..309N}. The loop geometry in the corona of HR 1099 has been studied in the past reporting loop length of the order of $10^{10}$ cm \citep[][]{2006ApJS..164..173M,2016PASJ...68...90T}. The abundance study of the corona of HR 1099 using high-resolution spectroscopy in X-ray domain has been performed by various authors in the past, showing the presence of the i-FIP effect in the corona of this star using various plasma models like {\sc CIE}, {\sc MEKAL}, {\sc APEC} etc. \citep[][]{2024MNRAS.528.4591B,2022A&A...659A...3S,2013ApJ...768..135H,2004ApJS..153..317O,2001A&A...365L...7D}.

This study delves into the X-ray flares on HR 1099, utilizing data obtained from the XMM-Newton satellite. By making a detailed study of XMM-Newton observations and its X-ray spectra and light curves, we studied three superflares and derived their physical properties, loop parameters, and abundance variations. The activity level of HR 1099 appears to have remained relatively stable over time, making this type of study valuable for enhancing our understanding of the star's physical properties. It also contributes to the long-term monitoring of superflares and provides insights into variations in their behavior.

The present paper is structured in the following manner: In Section \ref{sec:obs}, we provide details on the observations and data reduction processes. Section \ref{sec:results} depicts an in-depth explanation of our analysis procedures and presents the results of our X-ray temporal and spectral investigations. Lastly, in Section \ref{sec:discussion}, we present discussions concerning the results and conclusive findings.

\section{Observations and Data Reduction} \label{sec:obs} 
Observations of HR 1099 were conducted using state-of-the-art instruments aboard the XMM-Newton satellite, including the European Photon Imaging Cameras (EPIC), consisting of MOS and PN cameras and the Reflection Grating Spectrometers \citep[][]{2001A&A...365L..27T,2001A&A...365L..18S,2001A&A...365L...7D}. 

\begin{table*}
	\centering
	\caption{Log of observations of HR 1099.}
	\label{tab:log_table}
	\begin{tabular}{cclccccccc} 
		\hline\hline
		Set & Obs ID & Instruments & Start time & Exposure & Src Radius & Bkg Radius & Mode & Filter & Offset \\
		    & & & UTC & (s) & (") & (") & & & ($^\prime$)\\
		\hline
		 S1 & 0116150601 & PN & 29/01/2000 \enspace15:41:01 & 54234 & 55,15* & 40,40 & Full Frame & Thick & 0.145 \\ 
      S2 & 0791980501 & PN & 25/02/2020 \enspace12:15:40 & 10000 & 60 & 43,42 & Small Window & Thick & 0.145\\  
         &            & RGS & 25/02/2020 \enspace01:30:57 & 48917 & & & & & 0.145\\

		\hline
	\end{tabular}
    \\
     ~~~* These values correspond to the inner and outer radii of the chosen annular region
\end{table*}
HR 1099 has been observed 32 times by XMM-Newton between 2000 and 2024. After reviewing the PPS data products, we found that only 16 of these observations are suitable for analysis. Among these, six observations exhibited flares. Four of the flaring events have already been analyzed in previous studies.
The log of observations, which are analyzed in this paper, is given in Table \ref{tab:log_table}. We performed data reduction for the EPIC and RGS instruments using XMM-Newton's Science Analysis System (SAS) software version 21.0.0, along with the updated calibration files. To process raw EPIC data, we employed the tasks \textsc{emproc} and \textsc{epproc} for the MOS and PN data, respectively. To ensure data quality, we focused our analysis on the energy range of 0.3 to 10.0 keV for the EPIC data, as this range effectively minimizes high-energy background contributions. We used the task \textsc{evselect} to identify any intervals with high background proton flare, specifically for energy levels exceeding 10 keV. All datasets were found to be free from proton flare events. Furthermore, we assessed the presence of pile-up effects using the {\sc epatplot} task. Significant pile-up effects were observed in the dataset for observation of set S1, while no pile-up effect was detected in the case of set S2.

X-ray light curves and spectra were extracted for all PN observations by selecting circular regions centered on the source for all source counts. Background regions devoid of any source were chosen from the same CCD. The source and background radii for both the sets are mentioned in Table \ref{tab:log_table}. To mitigate the pile-up effects in observation data set S1, we opted for annulus regions with inner and outer radii of 15" and 55".

All X-ray light curves obtained from PN observations underwent corrections for high background contributions and other effects using the {\sc epiclccorr} task. Using the \textsc{especget} task we generated source and background spectra, along with auxiliary response files (ARF) and redistribution matrix files (RMF). We employed the \textsc{grppha} task to optimize the spectral analysis to ensure that all X-ray spectra were grouped into bins with a minimum of 20 counts each.

We conducted reduction procedures for the raw RGS data with the \textsc{rgsproc} task to produce event files and other spectral products. Subsequently, we used the \textsc{rgslccorr} and \textsc{rgcombine} tasks to generate combined light curves and spectra for RGS1 and RGS2, respectively. These grouped spectra were then utilized for subsequent analysis.
 \begin{figure*}
\centering

\subfigure[S1]{\includegraphics[width=0.93\columnwidth,trim={0.2cm 1.0cm 0.0cm 3.5cm}, clip]{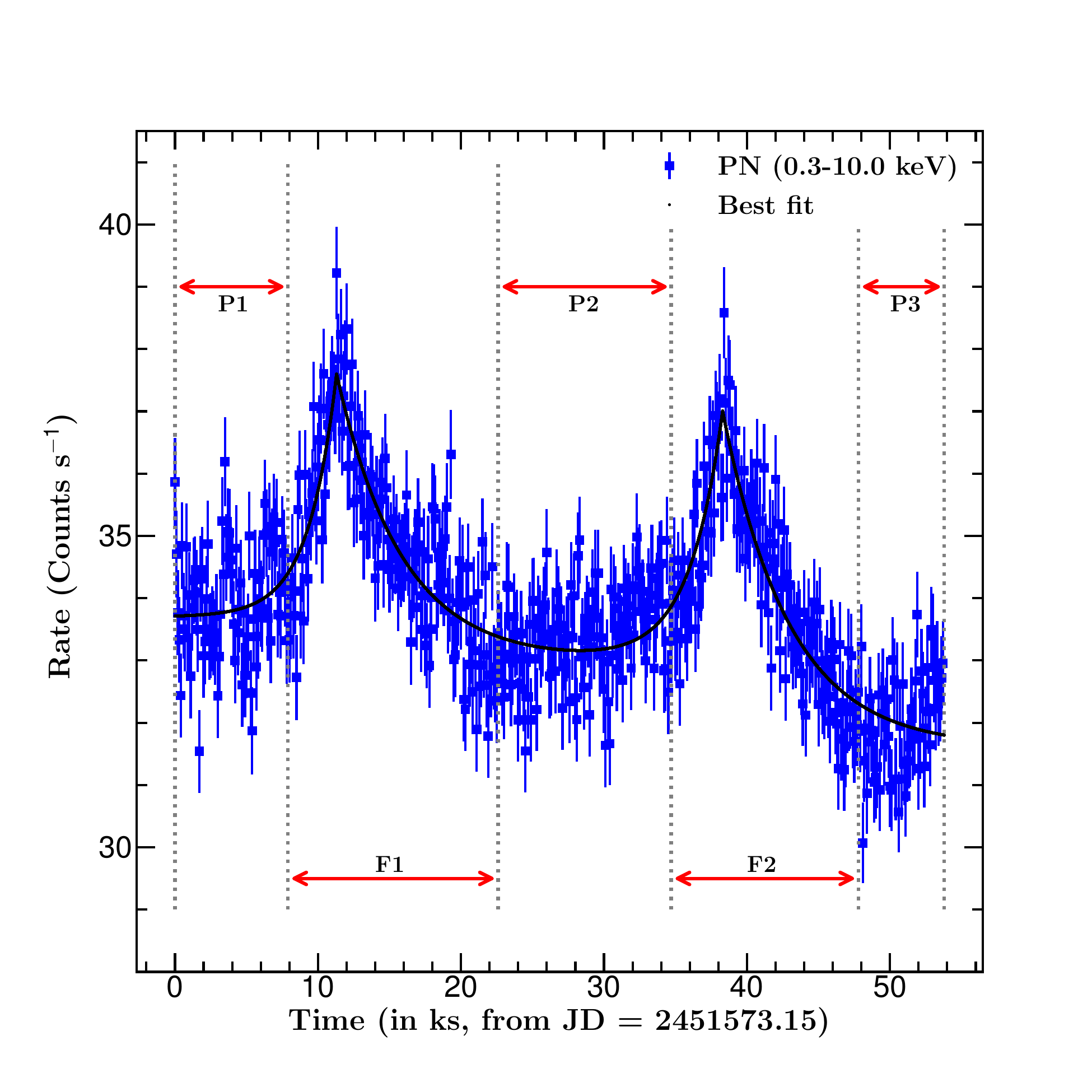}}
\subfigure[S2]{\includegraphics[width=0.93\columnwidth,trim={0.2cm 1.0cm 0.0cm 3.5cm}, clip]{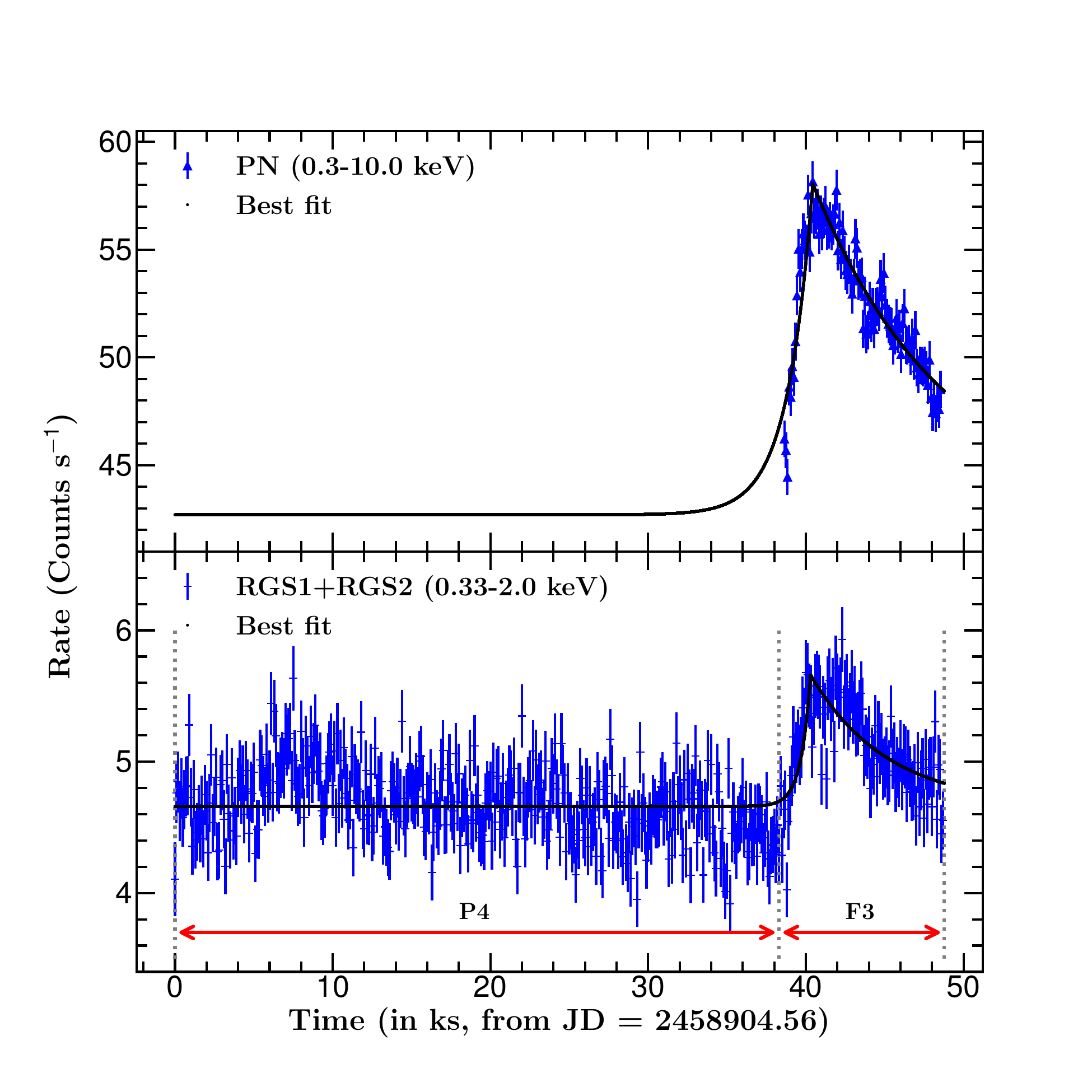}} 
\caption{X-ray light curves of HR 1099 in different detectors. The flaring regions are marked by the vertical lines.}
\label{fig:lcs_all}
\end{figure*}

\begin{figure*}
\centering
\subfigure[S1]{\includegraphics[width=0.93\columnwidth,trim={0.2cm 1.0cm 0.0cm 3.5cm}, clip]{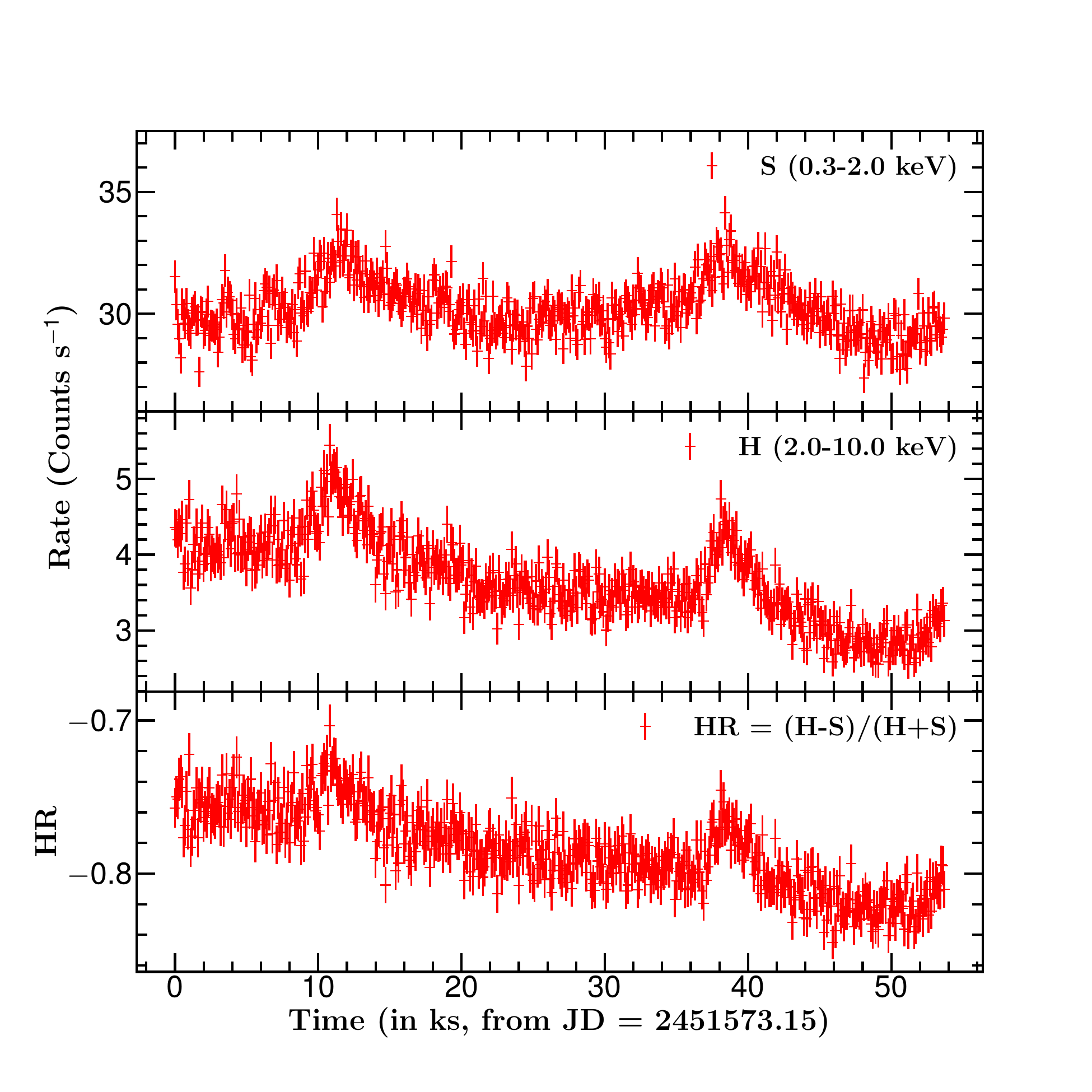}}
\subfigure[S2]{\includegraphics[width=0.93\columnwidth,trim={0.2cm 1.0cm 0.0cm 3.5cm}, clip]{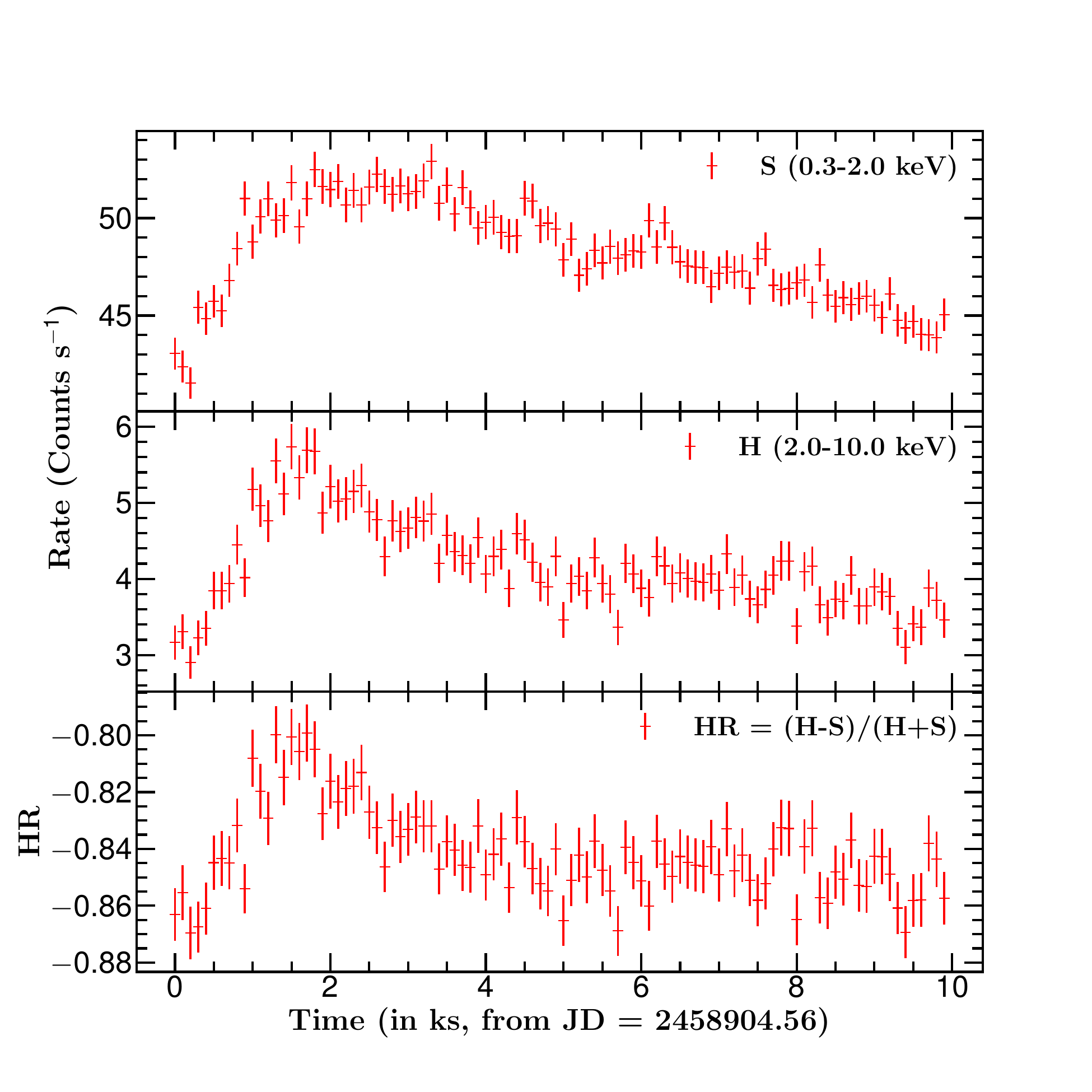}}
\caption{Soft and hard X-ray light curves along with the HR curve with 100s binning.}
\label{fig:lcs_HR}
\end{figure*}

\section{Analysis and Results}
\label{sec:results}
In this section, we focus on analyzing the temporal and spectral characteristics of the X-ray observations.

\subsection{X-ray Light Curves}
\label{sec:FLC}
Figure \ref{fig:lcs_all} represents the X-ray light curves with 100 s binning, obtained from PN and RGS detectors covering the energy range 0.3 - 10 keV and 0.33 - 2.0 keV, respectively. The temporal variability of counts exhibits flare-like features with a significant increase in the count rate observed during both observations. We identified a total of three flares, namely F1, F2, and F3. The pre and post-flare time segments with nearly constant counts are referred to as P1 to P4, serving as proxies for quiescent states. All these regions are marked in Figure \ref{fig:lcs_all}. We also used the criteria to rule out the possibility of false flare detection using only flares with a duration longer than 5 min and having at least 5 consecutive data points exceeding 3 $\sigma$ \citep[][]{2015EP&S...67...59M,2012Natur.485..478M}. However, we were unable to analyze pixel-level data as described in \cite{2012Natur.485..478M} due to the broad Point Spread Function (PSF) of XMM-Newton, with a Full Width at Half Maximum (FWHM) of approximately 15 arcseconds for the MOS cameras and around 6.6 for the PN \citep[][]{2001A&A...365L..27T,2001A&A...365L..18S}. Thus, the CCDs are primarily aimed at spectral and temporal analysis rather than spatially precise, pixel-level variability studies. Additionally, as noted in section \ref{sec:obs}, we confirmed the absence of high-energy background flares and other X-ray sources in the vicinity. The active K-type subgiant, being the primary X-ray emitter alongside a less active G-type main-sequence companion, rules out any flare contamination from other sources \citep[][]{1976AJ.....81..771B,1989A&A...211..173L}.

Additionally, we have plotted the light curve in two energy bands, namely soft (S; 0.2-2.0 keV) and hard (H; 2.0-10.0 keV) energy bands, along with the hardness ratio (HR), defined as HR = (H-S)/(H+S), in Figure \ref{fig:lcs_HR}. The HR curve closely follows the flaring events in the light curves, indicating the rise in temperature of the X-ray emitting region with an increase in HR during the flares. The HR increase confirms the coronal heating as well as it rules out the instrumental artifacts.

Following \cite{2024MNRAS.527.1705D}, we derived the rise and decay times for all three flares. The e-folding rise times for the flares were found to range from 1.6 to 2.3 ks, whereas the decay times ranged from 4.5 to 8.5 ks, indicating a rapid rise and slower decay pattern. Table \ref{tab:lc_fitting} lists the rise and decay times obtained from the light curves, and Figure \ref{fig:lcs_all} illustrates the modeled light curves. The peak flare count rates were increased by 16 to 36\% of the quiescent state count rate in these flares. 

\begin{table}
    \centering
    \caption{The best-fit parameters for all the observed flares.}  \label{tab:lc_fitting}
    \begin{tabular}{lccccc}
         \hline
         Parameters ($\rightarrow$)& $\tau_{r}$ & $\tau_{d}$ & A$_P$  & A$_Q$ \\
         Flare ($\downarrow$)  & (ks) & (ks) & (counts s$^{-1}$) &  (counts s$^{-1}$)\\        
         \hline
         S1/F1 (PN)& 2.0$\pm$0.2   & 4.5$\pm$0.2 & 39.2$\pm$0.7 & 33.7\\
         S1/F2 (PN)& 2.3$\pm$0.2   & 4.6$\pm$0.4 & 38.6$\pm$0.7 & 33.0\\
         S2/F3 (RGS) & 0.6$\pm$0.1 & 4.9$\pm$0.7 & 5.9$\pm$0.2 & 4.6\\
                 (PN)& 1.6$\pm$0.1   & 8.5$\pm$0.4 & 58.2$\pm$0.9 & 42.7*\\
          \hline 
    \end{tabular}
    ~~\\
    A$_P$ and A$_Q$ are the count rate of the flare peak and quiescent states, respectively. \\
    * Quiescent state count rates were estimated from RGS1 order 1 with an average count rate during the P4 state of 1.57 count s$^{-1}$  using the WebPIMMS\footnote{https://heasarc.gsfc.nasa.gov/cgi-bin/Tools/w3pimms/w3pimms.pl}.
   
\end{table}

\subsection{ EPIC Spectral Analysis} \label{sec:TRS}

To monitor variations in X-ray spectral parameters across all observations, we performed X-ray spectral analysis of quiescent and flaring states separately on the PN data, as it provides a better signal-to-noise (S/N) ratio. The spectral analysis was performed with Astrophysical Plasma Emission Code \citep[{\sc apec};][]{2001ApJ...556L..91S} using {\sc xspec}, version 12.12 \citep{1996ASPC..101...17A}. The detailed spectral analyses are described in forthcoming sections.

\subsubsection{The quiescent state spectroscopy} \label{sec:TRS_qui}

For determining the quiescent temperature and emission measure, pre/post-flare segments with nearly constant counts, marked as P1 to P4, were considered as the proxy of the quiescent. For set S1, we modeled the segmented spectra from P1 to P3 using one temperature (1-T), two temperatures (2-T), three temperatures (3-T), and four-temperature (4-T) plasma models {\sc apec} with varying parameters such as temperatures, corresponding normalization parameters, and abundances. Here, we used the {\sc tbabs} absorption model for hydrogen column density (\nh) and fixed it to the maximum value of galactic \nh of $10^{18}$ cm$^{-2}$ for HR 1099. We adopted solar abundances ($Z_{\odot}$) from \citet{2000ApJ...542..914W}. Abundances of all components were tied and varied together for the multi-component plasma model. The 3-T model fits the spectra well but gives varying temperatures for each P1, P2, and P3 segment. While adding one more component, the first three cool temperatures remained constant, whereas the fourth temperature varied for all the quiescent state segments. The average quiescent temperatures of cool components were found to be 0.26, 0.5, and 1.08 keV, with a weighted average temperature of 0.82 keV, calculated using the following equation.
\begin{equation}\label{eqn:avg}
T_{Q} = \frac{\sum_{i=1}^3 T_{i} EM_{i}}{\sum_{i=1}^3 EM_{i}} ; \quad EM_{Q} = \frac{1}{3}\sum_{i=1}^3 EM_{i}, 
\end{equation}

For set S2, the PN data included only the flare part without the pre/post-flare segments throughout the observations. Therefore, we divided flare F3 into segments (rise, peak, and decay) with sufficient counts and applied the same procedure as applied for set S1. Again, the fourth temperature was found to be varying, while the first three temperatures remained nearly constant across all the segments. The coolest three temperatures were found to be similar to that for set S1 within one sigma level. Thus, for further analysis, the quiescent corona of HR 1099 was assumed to be best represented by a three-temperature plasma. All the derived spectral parameters for the quiescent states are given in Table \ref{tab:qui_3a_all}.

\begin{table*}
    \centering 
    {\setlength{\tabcolsep}{3.4pt}
    \caption{Spectral parameters of the quiescent using 4-T APEC model for all the data sets, within a 68\% confidence range.}
    \begin{tabular}{cccccccccccccc} 
         \hline
         Para ($\rightarrow$)& $kT_{1}$ & $kT_{2}$ & $kT_{3}$ & $kT_{4}$ & $T_{Q}$* && $EM_{1}$ & $EM_{2}$ & $EM_{3}$ & $EM_{4}$ & $EM_{Q}$* & Z  & $\chi_\nu^2 $ (dof) \\   \cline{2-6} \cline{8-12}
         Set ($\downarrow$) & \multicolumn{5}{c}{(keV)} &&  \multicolumn{5}{c}{($10^{53} cm^{-3}$)}   & (Z$_\odot$) &    \\ 
         \hline
         S1  & 0.26$_{-0.03}^{+0.03}$ & 0.5$_{-0.1}^{+0.1}$ & 1.08$_{-0.02}^{+0.03}$ & 2.3$_{-0.09}^{+0.1}$ & 0.82$_{-0.05}^{+0.06}$ && 1.4$_{-0.5}^{+0.4}$ & 0.8$_{-0.4}^{+0.5}$ & 3.9$_{-0.3}^{+0.4}$ & 5.9$_{-0.3}^{+0.3}$ & 2.0$_{-0.2}^{+0.3}$ & 0.26$_{-0.03}^{+0.03}$ & 1.09 (566)\\
         S2 & 0.26$_{-0.02}^{+0.02}$ & 0.59$_{-0.08}^{+0.09}$ & 1.07$_{-0.03}^{+0.04}$ & 2.6$_{-0.1}^{+0.2}$ & 0.8$_{-0.1}^{+0.1}$ && 29$_{-4}^{+3}$ & 15$_{-4}^{+5}$ & 70$_{-7}^{+7}$ & 64$_{-6}^{+6}$ & 38$_{-3}^{+3}$ & 0.34$_{-0.03}^{+0.03}$ & 1.14(502)\\
         \hline
    \end{tabular}
    \\ \label{tab:qui_3a_all}
    ~~~*For the calculation of $T_{Q}$ and $EM_{Q}$, only the first three temperatures and emission measures have been used.\\
    }
    
\end{table*}

\subsubsection{Spectral evolution during the flares} \label{sec:TRS_flare}

We divided the flaring part of the light curve into distinct time intervals of rise, peak, and decay phases to understand the flare evolution. These phases are denoted as Ri, P, and Di, where the index 'i' can take values from 1 onwards. The duration of each flare segment was chosen to ensure an equal number of counts within each interval.

\begin{figure}
\centering

\subfigure[S1]{\includegraphics[width=0.99\columnwidth,trim={0.5cm 1.7cm 2.5cm 3.5cm}, clip]{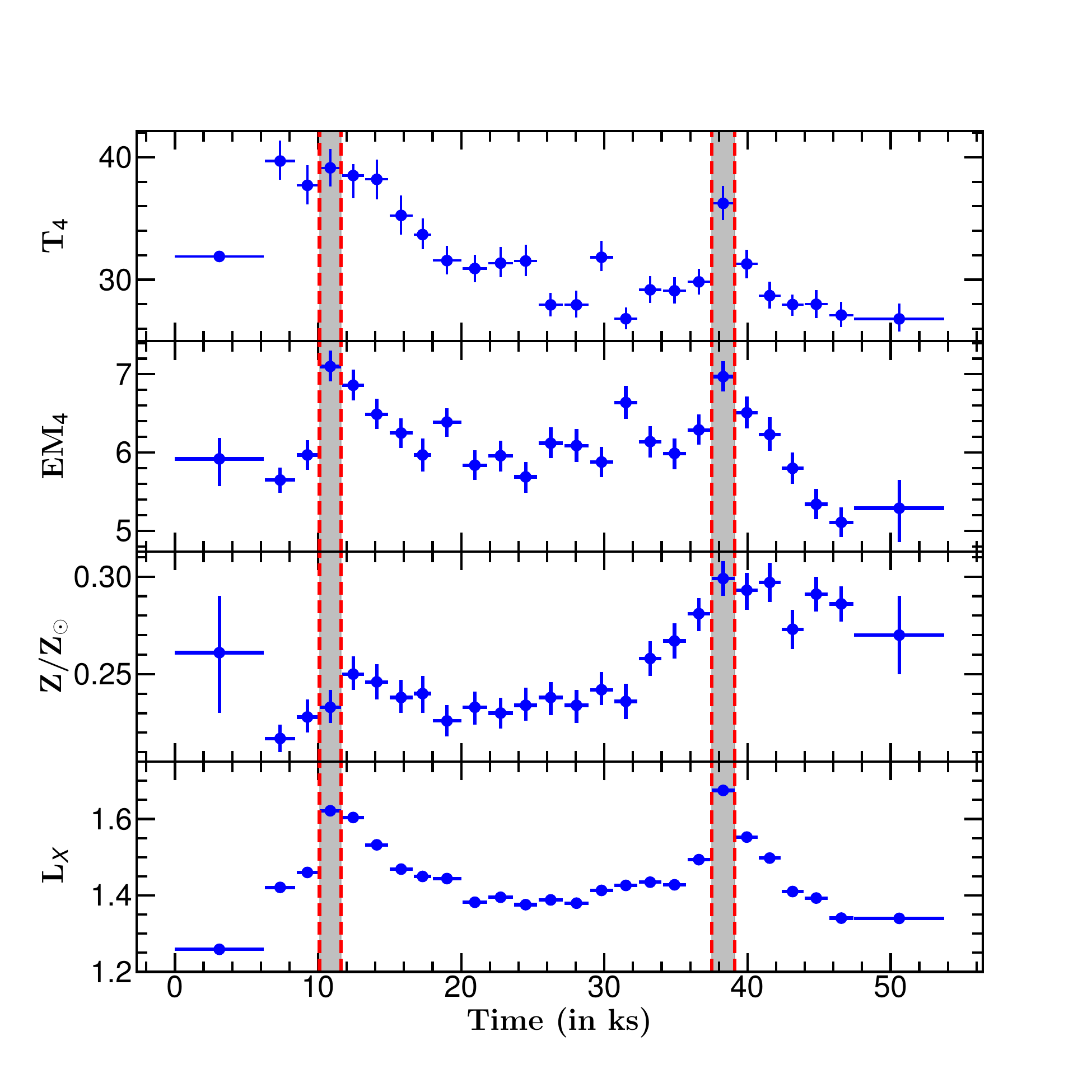}}
\subfigure[S2]{\includegraphics[width=0.99\columnwidth,trim={0.4cm 1.7cm 2.5cm 3.5cm}, clip]{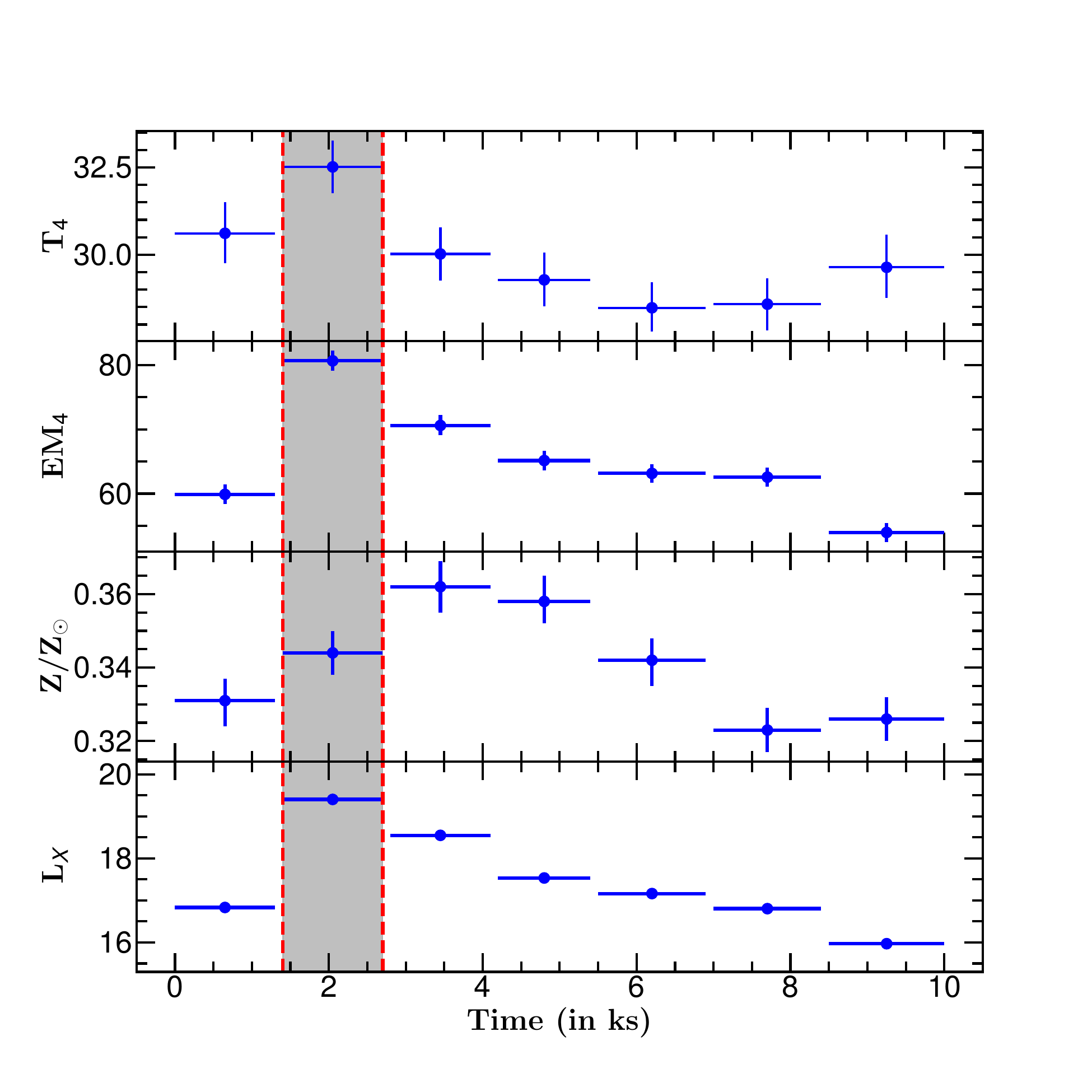}}
\caption{The temporal variation of spectral parameters $T_{4}$, $EM_{4}$, Z, and Luminosity \lxf for sets S1 and S2. The $T_{4}$ is in units of MK, $EM_{4}$ in $10^{53}$ $cm^{-3}$, \lxf in $10^{31}$ erg/s. The shaded regions show the peak phase of the corresponding flare.}
\label{fig:tvspara}
\end{figure}

We applied a consistent approach for all flares analyzed here. Each segmented spectrum was modeled using a 4-T {\sc apec} model. The temperatures and emission measures of the first three plasma components were fixed at their corresponding quiescent values, allowing the parameters of the fourth component to vary in order to obtain the flare-specific parameters. The parameters derived for each flare are given in Table \ref{tab:flare_apec_all}. 
The temporal variations of these parameters are illustrated in Figure \ref{fig:tvspara}, demonstrating that nearly all parameters peak during the peak phase of the flare. The peak flare temperature, $kT_4$, was found to be 3.4, 3.1, and 2.8 keV, while the peak emission measure, $EM_4$, reached 7.1, 7.0, and $81 \times 10^{53} cm^{-3}$ for flares F1, F2, and F3, respectively. We determined the unabsorbed flux for each flare segment using the convolution model {\sc cflux} and integrated this flux across all segments to obtain the total flux (F). From this, we derived the luminosity (\lxf = 4$\pi$$D^2$F) and the total flare energy ($E_{X,Total}$ = \lxf $\times$ ($\tau_{r}$+$\tau_{d}$)), where D represents the distance to the object in cm. The resulting values for \lxf and $E_{X,Total}$ are presented in Table \ref{tab:final_para}. The peak abundance (Z) and the peak X-ray luminosity (\lxf) of flares were found to be  0.25, 0.299, and 0.362 Z$_\odot$ and 10$^{31.21}$, 10$^{31.23}$, and 10$^{32.29}$ erg s$^{-1}$ for the flare F1, F2, and F3, respectively. We found that temperature peaked either earlier than the emission measure or simultaneously with the emission measure. The abundance peaked after the emission measure for flares F1 and F3, whereas for flare F2, it peaked with the emission measure and remained constant thereafter for the initial phase of the flare decay.

\begin{table*}
    \centering
    \caption{Best fit spectral parameters for each time segment during the flares F1 to F3 within 68\% confidence interval.}
    \begin{tabular}{ccccccc}
         \hline
         Parameters ($\rightarrow$) & Flare& $kT_{4}$ & $EM_{4}$ & Z  & \lxf  & $\chi_\nu^2 $ (dof) \\
         Flare ($\downarrow$) & Segments& (keV) & ($10^{53} cm^{-3}$) & (Z$_\odot$) &  ($10^{31}$ erg s$^{-1}$)  \\
         \hline 
        F1  & R1     & 3.4$_{-0.1}^{+0.1}$    & 5.7$_{-0.2}^{+0.2}$   & 0.217$_{-0.007}^{+0.007}$     & 1.421$_{-0.009}^{+0.009}$   & 1.12 (416) \\ 
		  & R2     & 3.3$_{-0.1}^{+0.1}$    & 6.0$_{-0.2}^{+0.2}$   & 0.228$_{-0.009}^{+0.009}$     & 1.46$_{-0.01}^{+0.01}$      & 1.08 (372) \\ 
		& P      & 3.4$_{-0.1}^{+0.1}$    & 7.1$_{-0.2}^{+0.2}$   & 0.233$_{-0.009}^{+0.009}$     & 1.62$_{-0.01}^{+0.01}$      & 1.16 (395) \\ 
		& D1     & 3.32$_{-0.09}^{+0.1}$  & 6.9$_{-0.2}^{+0.2}$   & 0.250$_{-0.009}^{+0.009}$     & 1.60$_{-0.01}^{+0.01}$      & 1.14 (386) \\ 
		& D2     & 3.29$_{-0.08}^{+0.09}$ & 6.5$_{-0.2}^{+0.2}$   & 0.246$_{-0.009}^{+0.009}$     & 1.53$_{-0.01}^{+0.01}$      & 1.02 (388) \\ 
		& D3     & 3.0$_{-0.1}^{+0.1}$    & 6.2$_{-0.2}^{+0.2}$   & 0.238$_{-0.008}^{+0.009}$     & 1.47$_{-0.01}^{+0.01}$      & 1.04 (377) \\ 
		& D4     & 2.9$_{-0.1}^{+0.1}$    & 6.0$_{-0.2}^{+0.2}$   & 0.24$_{-0.01}^{+0.01}$        & 1.45$_{-0.01}^{+0.01}$      & 1.05 (346) \\ 
       F2   & R1     & 2.51$_{-0.09}^{+0.09}$ & 6.0$_{-0.2}^{+0.2}$   & 0.267$_{-0.009}^{+0.009}$     & 1.43$_{-0.01}^{+0.01}$      & 1.0 (371) \\ 
		& R2     & 2.57$_{-0.09}^{+0.09}$ & 6.3$_{-0.2}^{+0.2}$   & 0.281$_{-0.009}^{+0.009}$     & 1.49$_{-0.01}^{+0.01}$      & 0.97 (375) \\ 
		& P      & 3.1$_{-0.1}^{+0.1}$    & 7.0$_{-0.2}^{+0.2}$   & 0.299$_{-0.009}^{+0.009}$     & 1.68$_{-0.01}^{+0.01}$      & 1.05 (405) \\ 
		& D1     & 2.7$_{-0.1}^{+0.1}$    & 6.5$_{-0.2}^{+0.2}$   & 0.293$_{-0.01}^{+0.009}$      & 1.55$_{-0.01}^{+0.01}$      & 1.08 (372) \\ 
		& D2     & 2.5$_{-0.1}^{+0.1}$    & 6.2$_{-0.2}^{+0.2}$   & 0.297$_{-0.009}^{+0.01}$      & 1.50$_{-0.01}^{+0.01}$      & 0.96 (362) \\ 
		& D3     & 2.41$_{-0.08}^{+0.08}$ & 5.8$_{-0.2}^{+0.2}$   & 0.27$_{-0.01}^{+0.01}$        & 1.41$_{-0.01}^{+0.01}$      & 1.04 (357) \\ 
		& D4     & 2.4$_{-0.1}^{+0.1}$    & 5.3$_{-0.2}^{+0.2}$   & 0.291$_{-0.009}^{+0.009}$     & 1.39$_{-0.01}^{+0.01}$      & 1.15 (359) \\ 
		& D5     & 2.3$_{-0.1}^{+0.1}$    & 5.1$_{-0.2}^{+0.2}$   & 0.286$_{-0.009}^{+0.009}$     & 1.34$_{-0.01}^{+0.01}$     & 0.93 (363) \\
        F3  & R      & 2.64$_{-0.07}^{+0.08}$ & 60$_{-1}^{+2}$        & 0.331$_{-0.006}^{+0.006}$     & 16.83$_{-0.01}^{+0.01}$     & 1.17 (489) \\ 
	    & P      & 2.80$_{-0.06}^{+0.07}$ & 81$_{-2}^{+2}$        & 0.344$_{-0.006}^{+0.006}$     & 19.40$_{-0.01}^{+0.01}$     & 1.09 (544) \\ 
	    & D1     & 2.59$_{-0.07}^{+0.07}$ & 71$_{-2}^{+2}$        & 0.362$_{-0.007}^{+0.007}$     & 18.55$_{-0.01}^{+0.01}$     & 1.13 (519) \\ 
	    & D2     & 2.52$_{-0.06}^{+0.07}$ & 65$_{-2}^{+2}$        & 0.358$_{-0.007}^{+0.007}$     & 17.53$_{-0.01}^{+0.01}$     & 1.17 (495) \\ 
	    & D3     & 2.45$_{-0.06}^{+0.06}$ & 63$_{-1}^{+2}$        & 0.342$_{-0.006}^{+0.006}$     & 17.16$_{-0.01}^{+0.01}$     & 1.12 (509) \\ 
	      & D4     & 2.46$_{-0.06}^{+0.06}$ & 63$_{-1}^{+2}$        & 0.323$_{-0.006}^{+0.006}$     & 16.80$_{-0.01}^{+0.01}$     & 1.12 (509) \\
   \hline
    \end{tabular}
    \label{tab:flare_apec_all}
\end{table*}

\subsection{RGS Spectral Analysis}
\label{sec:RGS_spectra}

The RGS1 and RGS2 spectra for set S2 were generated for both the quiescent state (P4) and the flaring state (F3) and fitted with the {\sc vapec} model. Additionally, the {\sc tbabs} model was used to account for the \nh, which was fixed to a minimum value of $10^{18}$ cm$^{-2}$. All temperatures and corresponding emission measures were left as free parameters for the quiescent state spectral fitting. The abundances of He, Ni, and Al were set to the solar photospheric values, while the other abundances were allowed to vary and were tied among each temperature component. We found that a 3-T plasma model provided the best fit for the P4 quiescent segment. 

For the flaring spectra,  a 4-T {\sc vapec} model was used for the fitting. The first three temperatures and corresponding normalization parameters were fixed at the quiescent values as obtained from spectral fitting of the P4 region, and all abundances were tied among the different components. Table \ref{tab:vapec_all} represents the best-fit model parameters for both the quiescent and flaring phases within a 68\% confidence interval.

\begin{figure}
\centering

\includegraphics[width=0.54\textwidth,trim={0.4cm 0.0cm 1.5cm 2.0cm}, clip]{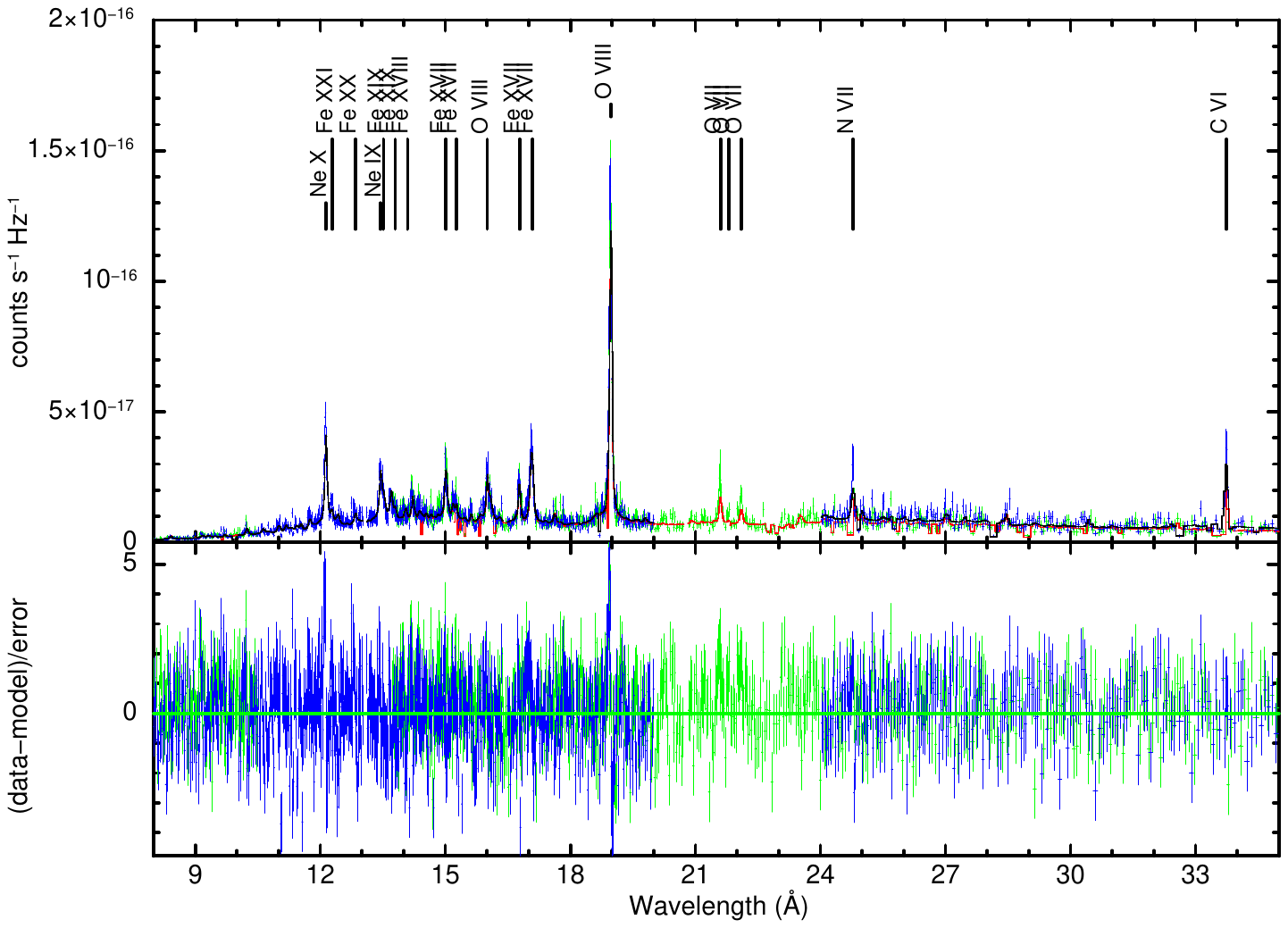}
\includegraphics[width=0.54\textwidth,trim={0.4cm 0.0cm 1.5cm 2.0cm}, clip]{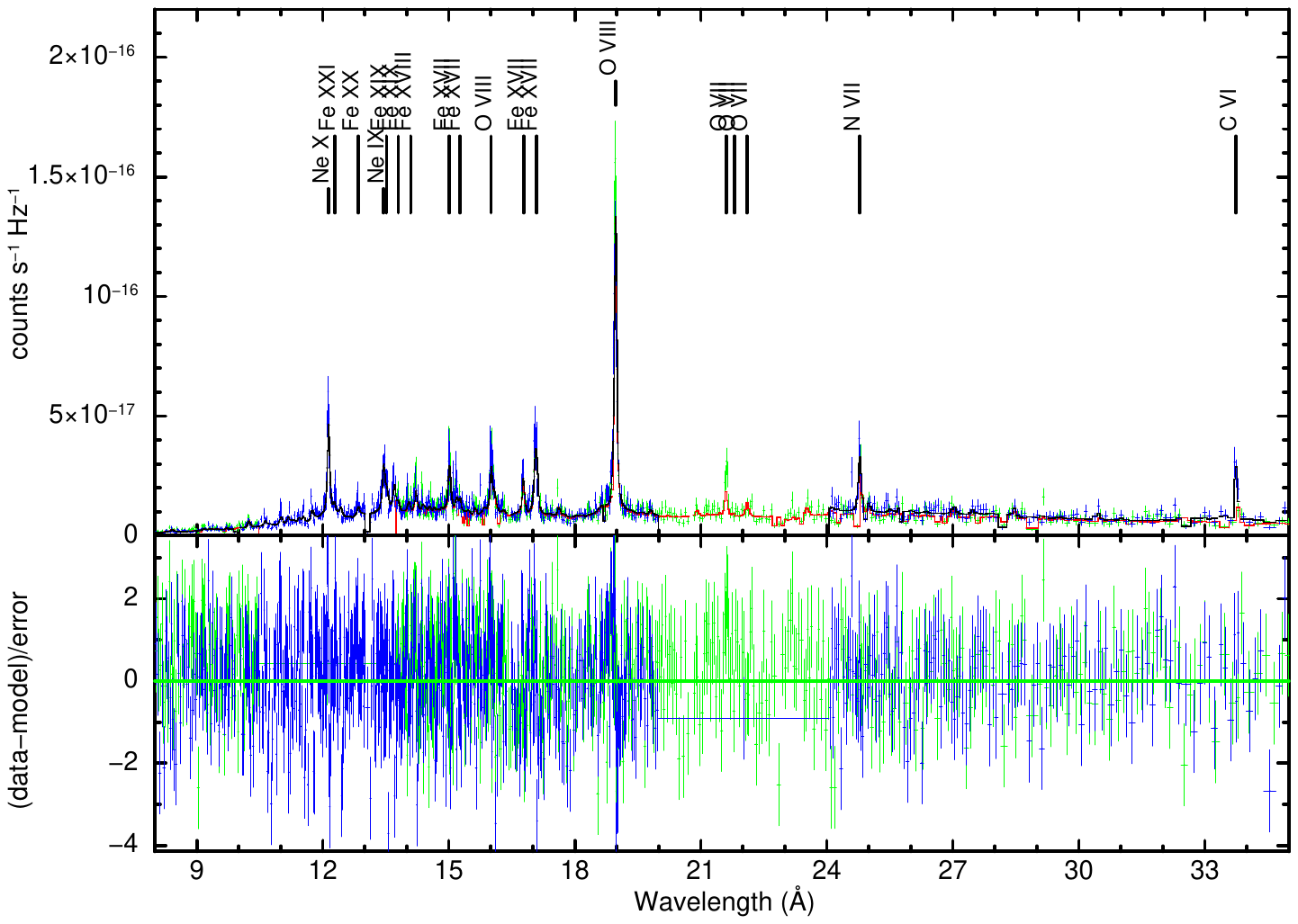}
\caption{The RGS1 and RGS2 spectra of pre-flare state P4 and flare F3 are shown in the upper panel with 3-T {\sc vapec} and 4-T {\sc vapec} models, respectively. The lower panel shows the residual of the modeled spectra.}
\label{fig:rgs_spectra}
\end{figure}

In Figure \ref{fig:rgs_spectra}, we present the RGS1 and RGS2 spectra of HR 1099 for flare F3 and its pre-flare state P4, including emission lines from various elements. The best-fit thermal plasma model, {\sc vapec}, is illustrated with solid lines for both the flare and quiescent states. The residuals of the best-fit model are shown in the lower panel. 

\begin{table}
    \centering
    {\setlength{\tabcolsep}{10pt}
    \caption{The best-fit optimal spectral parameters derived from fitting the RGS spectra within a 68\% confidence level.}     \label{tab:vapec_all}
    \begin{tabular}{lcc}
         \hline\hline
         Segments ($\rightarrow$) & P4 & F3  \\
         Parameters ($\downarrow$) &  Z/$Z_\odot$ & Z/$Z_\odot$ \\
         \hline
         $kT_{1}$ (keV)                & 0.39$_{-0.02}^{+0.01}$  &       ....                    \\
         $kT_{2}$ (keV)                & 0.83$_{-0.02}^{+0.01}$  &       ....                    \\
         $kT_{3}$ (keV)                & 1.74$_{-0.05}^{+0.05}$  &       ....                        \\
         $kT_{4}$ (keV)                &      ....               &  1.9$_{-0.4}^{+1.4}$       \\
         $EM_{1}$ ($10^{53}$cm$^{-3}$) & 1.02$_{-0.09}^{+0.06}$  &      ....                         \\
         $EM_{2}$ ($10^{53}$cm$^{-3}$) & 2.27$_{-0.09}^{+0.09}$  &      ....                         \\
         $EM_{3}$ ($10^{53}$cm$^{-3}$) & 4.6$_{-0.1}^{+0.1}$     &      ....                         \\
         $EM_{4}$ ($10^{53}$cm$^{-3}$) &       ....              &   1.08$_{-0.03}^{+0.05}$      \\
         C              & 1.0$_{-0.1}^{+0.1}$     & 1.1$_{-0.1}^{+0.1}$    \\
         N              & 0.55$_{-0.09}^{+0.09}$  & 0.95$_{-0.07}^{+0.07}$   \\
         O              & 0.72$_{-0.03}^{+0.03}$  & 0.79$_{-0.02}^{+0.01}$ \\
         Ne             & 1.50$_{-0.07}^{+0.08}$  & 1.63$_{-0.07}^{+0.07}$  \\
         Mg             & 0.44$_{-0.05}^{+0.05}$  & 0.48$_{-0.04}^{+0.04}$   \\
         Si             & 0.70$_{-0.07}^{+0.07}$  & 0.86$_{-0.07}^{+0.07}$   \\
         S              & 0.38$_{-0.09}^{+0.1}$   & 0.49$_{-0.09}^{+0.09}$ \\
         Ar             & 0.9$_{-0.2}^{+0.2}$     & 1.3$_{-0.2}^{+0.2}$    \\
         Fe             & 0.26$_{-0.01}^{+0.01}$  & 0.28$_{-0.04}^{+0.04}$ \\
         $\chi_\nu^2$   & 1.41                 &  1.19                         \\
         (dof)          & 2502                 &  1249                         \\
         \hline
         \end{tabular}
         ~~\\
         \textbf{Note}. All the elemental abundances are in terms of solar photospheric abundances taken from \cite{2000ApJ...542..914W}.\\
         }
\end{table}

\subsubsection{FIP and inverse-FIP effect}
\label{sec:IFIP}
Due to fractionation processes associated with the first ionization potential (FIP), the elemental abundances in the corona differ from those in the underlying photosphere. In less active slow rotators like the Sun, elements with FIP < 10 eV are generally enhanced compared to those with FIP > 10 eV, a phenomenon termed the FIP effect \citep[][]{1992PhyS...46..202F,1995ApJ...443..416L,2000PhyS...61..222F}. On the other hand, fast-rotating stars, which exhibit higher magnetic activity, often show an inverse-FIP (i-FIP) effect \citep[see][]{2015LRSP...12....2L,2021ApJ...909...17L}. Some stars with magnetically inactive chromospheres show no detectable FIP bias \citep[][]{1994AAS...184.0522D}.

The measured abundances as a function of FIP, for both quiescent and flaring phases are shown in Figure \ref{fig:fip_all}. The elements with FIP < 10 eV like Mg, Fe, and Si were found to be under-abundant relative to the solar photospheric values in both states, with values of approximately $\sim$0.44, $\sim$0.26, and $\sim$0.70 for the quiescent state, and $\sim$0.48, $\sim$0.28, and $\sim$0.86 for the flaring state, respectively. Additionally, the abundances of N, O, and Si are lower in the quiescent state compared to the flaring state, while the remaining abundances are similar to the flaring values within a 68\% confidence interval. Thus, the evidence of the i-FIP effect is observed in HR 1099 in the case of both the quiescent and flaring states in the present observational baseline.

\subsection{Loop Modelling}
\label{sec:LLC}
\begin{figure}
\centering

\subfigure[]{\includegraphics[width=0.93\columnwidth,trim={0.5cm 0.0cm 1.0cm 2.0cm}, clip]{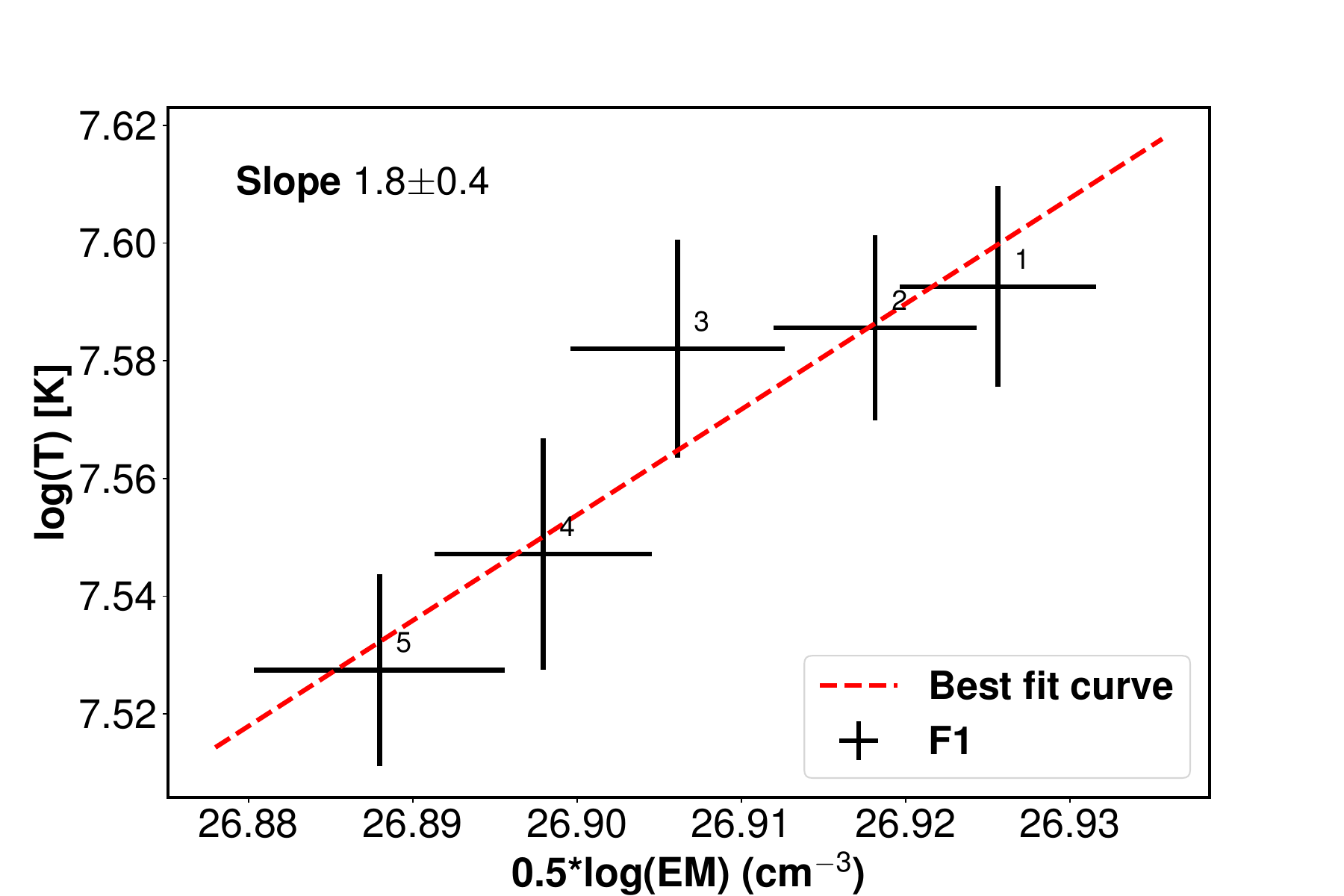}}
\subfigure[]{\includegraphics[width=0.93\columnwidth,trim={0.4cm 0.0cm 1.0cm 2.0cm}, clip]{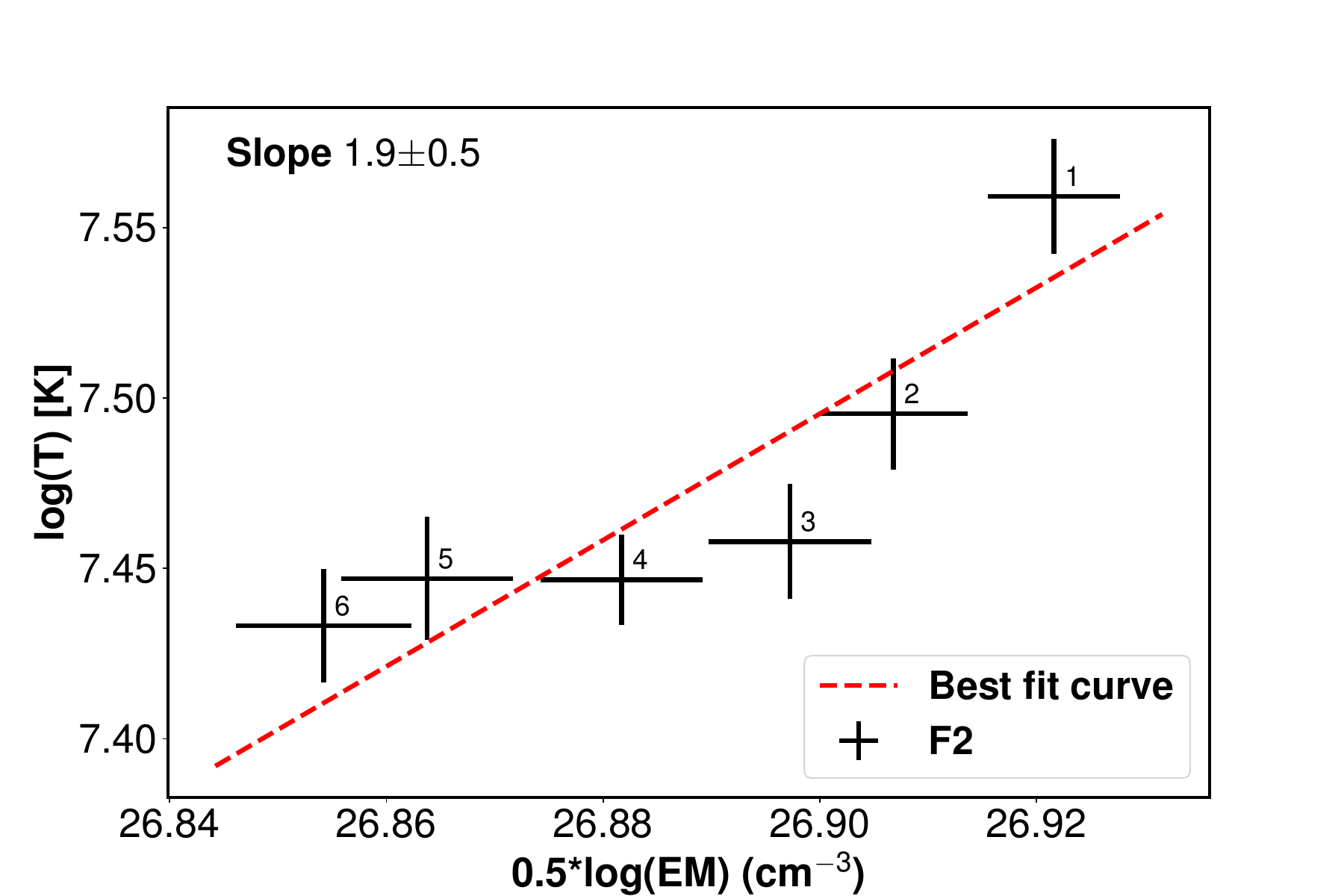}}
\subfigure[]{\includegraphics[width=0.93\columnwidth,trim={0.5cm 0.0cm 1.0cm 2.0cm}, clip]{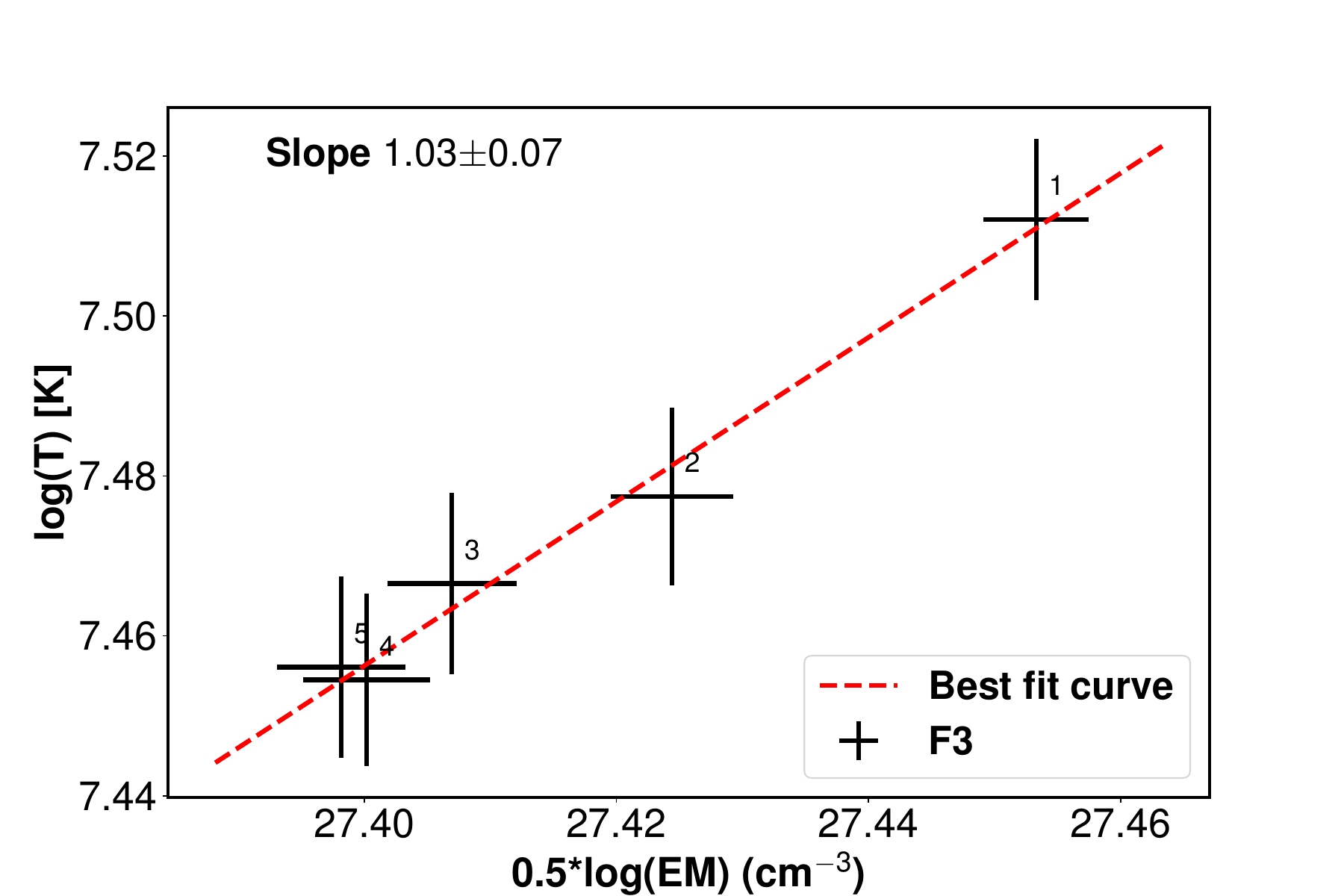}}
\caption{The density vs temperature (n-T) plot, where $EM^{1/2}$ is considered as the proxy of density. $\zeta$ is the slope of the n-T diagram from the decay phase of the flares (a) F1, (b) F2, and (c) F3.}
\label{fig:slope}
\end{figure}

Loop length is an essential parameter for determining the physical size of flaring loops and the star’s corona. Although stellar flares are spatially unresolved, analogies with the solar flares and loop models allow us to infer the coronal geometry. During a heating event, the temperature of the loop increases and peaks at temperature $T_{0}$, while chromospheric evaporation causes an increase in the density. Throughout the decay phase, the thermal conduction cooling becomes dominant, leading to a temperature drop during this stage of the flare. This cooling phase provides valuable insights into the energy release and transport mechanisms within the stellar corona.
Using the hydrodynamic loop model, \citet{1997A&A...325..782R} derived a relation for the semi-loop length of the flaring loops. This model assumes a dominant single coronal loop and incorporates both plasma cooling and heating effects during the flare's decay phase. The hydrodynamic model helps to understand the balance between conductive and radiative cooling processes. The semi-loop length ($L$) is given by \citet{1997A&A...325..782R}:

\begin{equation}
{\rm L = 2.7\times10^3\frac{\tau_{d}T_{max}^{1/2}}{F(\zeta)}  ~~cm~~
~~ for ~~~0.35 < \zeta \leq 1.6}
\end{equation}

\begin{equation}
   {\rm  with  \enspace F(\zeta) = \frac{0.51}{\zeta-0.35}+1.36}, \enspace and 
    \quad T_{max} = 0.13 T_{0}^{1.16}
    \label{eq:loop}
\end{equation}
Here, $\tau_d$ is the decay time in s, $F(\zeta)$ is the correction factor accounting for the heating, and $\zeta$ is the slope of the best fit straight line of log($\sqrt(EM)$) vs. log(T) diagram during the decay phase of the flare. Figure \ref{fig:slope} shows log($\sqrt(EM)$) vs. log(T) diagram (equivalent to the density vs temperature diagram) for all three flares along with the best fit straight line and corresponding $\zeta$ values. The value of $\zeta$ in the range of 0.36 to 1.6 suggests sustained heating during the flare's slow decay phase \citep[see][for detail]{2007A&A...471..271R}. For flare F1 and F2, the value of $\zeta$ is above the critical value but well with a 1$\sigma$ level, whereas for flare F3, $\zeta$ was found to be $1.03\pm0.07$. These values indicate the non-negligible heating during the decay phase of the observed flares.

The $T_{0}$ is the maximum best-fit segmented average temperature derived from the spectral fitting in units of K, whereas ($T_{max}$) is the maximum temperature of the flare calculated using the equation \ref{eq:loop}. The $T_{max}$ for the flare F1, F2, and F3 were found to be 84$\pm$3, 76$\pm$3, and 67$\pm$2 MK, respectively. 

The derived loop lengths were found to be in the range 5.9 -- 8.9 $\times 10^{10}$ cm as given in Table \ref{tab:final_para}.
Assuming a single cylindrical loop with half loop length $L$, we have derived some more physical parameters like loop volume (V), plasma density ($n_e$), plasma pressure at loop apex (P), the minimum magnetic field required to confine the plasma inside the flaring loop ($B_{min}$), total magnetic field ($B_{Total}$), heating rate ($E_{HR}$), and the total energy related to the heating rate ($E_{H,Total}$). \cite{2024MNRAS.527.1705D} have also performed such estimations for the corona of AB Dor. All these parameters are mentioned in Table \ref{tab:final_para}. 

\begin{table}
\centering
    \caption{Loop parameters.}
    \label{tab:final_para}
    \begin{tabular}{lccc}
    \hline
     \hline
Flare ($\rightarrow$)                   & F1               & F2                 & F3  \\  
Parameters ($\downarrow$)               &                  &                    &      \\ 
\hline
\lxf ($10^{31}$ \lum)                   & 10.55 $\pm$ 0.03 &   11.79 $\pm$ 0.03 &   106.3 $\pm$ 0.2\\
$E_{X,Total}$ ($10^{35}$ erg)           &  6.8 $\pm$ 0.3   &   8.1 $\pm$ 0.5    &   107 $\pm$ 4\\
T$_0$ (MK)                              & 39 $\pm$ 1       &   36 $\pm$ 1       &   32.5 $\pm$ 0.8  \\
T$_{max}$ (MK)                          & 84 $\pm$ 3       &   76 $\pm$ 3       &   67 $\pm$ 2          \\
L ($10^{10}$ cm)                        & > 6.0             &  >5.9              & 8.9 $\pm$ 0.5 \\
h/$R_{*}$                               & >0.14            &  >0.14             & 0.21$\pm$0.01 \\
V ($10^{31}$ \vol)                      & >1.36            & >1.29              & 4.4$\pm$0.8 \\ 
$n_{e}$ ($10^{11}$ \density)            & <2.5             & <2.6               & 4.8$\pm$0.4\\ 
P  ($10^3$ \pressure)                   & <5.3             & <4.9               & 8.0$\pm$0.7\\ 
$B_{min}$ (G)                           & <364             & <351               & 448$\pm$20\\ 
$B_{Total}$ (G)                         & $\sim$608        & $\sim$558          & 530$\pm$28\\ 
E$_{HR}$ ($10^{31}$ \lum)               & $\sim$2.0        & $\sim$1.4          & 1.4$\pm$0.2\\ 
$E_{H,Total}$ ($10^{35}$ erg)           & $\sim$1.3        & $\sim$1.0          & 1.4$\pm$2\\ 
M$_{CME}$ (10$^{19}$ g)                 & $\sim$3.04       & $\sim$3.25         & $\sim$14.5\\
\hline
    \end{tabular}
    ~~\\
    \textbf{Note.} Here,  loop height, h = 2L/$\pi$ and $R_{*}$ is the radius of the star. \\
\end{table}

\section{Discussion and conclusions}
\label{sec:discussion}

Our study of the active RS CVn binary HR 1099 provides a detailed analysis of the three energetic X-ray flares. The duration of these flares is found in the range of 2.8 to 4.1 h. The flare duration of these three flares is in between the flare duration observed in other flares in HR 1099, which were in the range of 1 to 9 hr \citep[][]{2012MNRAS.419.1219P, 2001A&A...365L.318A, 2007A&A...464..309N, 2004ApJS..153..317O}. However, these flares are of very short duration in comparison to the longest duration flares observed in HR 1099 on two different occasions,  with a flare duration of >2.3 and $\sim$ 2.8 days, respectively \citep[][]{2024AAS...24344706K, 2000PhyS...61..222F}.

Our analysis further explores the e-folding rise and decay times of these flares, revealing the rise time ranging from 10 to 38 minutes and decay time from 1.25 to 2.4 h. This pattern of a rapid rise followed by a more gradual decay aligns with established characteristics observed in other stellar flares from HR 1099 and similar RS CVn systems. For instance, previous studies have reported a rise time between 18 and 84 minutes and a decay time from 0.8 to 1.9 h \citep[][]{2012MNRAS.419.1219P, 2021MNRAS.505L..79Y}. Moreover, \cite{2004ApJS..153..317O} found the rise and decay times of the X-ray flares from HR 1099 in the range of 1.4 - 11.1 h and 2.6 - 13.2 h, respectively, for the observations from  ASCA (0.6 - 10 keV), RXTE (2 - 12 keV), and BeppoSAX (0.6 - 10 keV). Further, \cite{2006ApJS..164..173M} and \cite{2016PASJ...68...90T} reported even longer decay times, ranging from 1.6 to 6.6 h with EUVE (7-76 nm) and 1.7 to 18.6 h with MAXI/GSC (2-30 keV), respectively.
Additionally, the peak flare to quiescent state count rate ratio ($A_{P}$/$A_{Q}$) was found to be 1.2 for flares F1 and F2 and 1.4 for flare F3, which is the typical range for the energetic flares in the active stars \citep[][]{2008MNRAS.387.1627P,2012MNRAS.419.1219P,2024MNRAS.527.1705D}.  

We have also estimated the flare frequency based on observations from 1978 to 2024. During a total observing time of 142 days, 57 flares were detected, leading to an estimated flare frequency of 0.4 flares per day. This translates to approximately one flare per rotation period of HR 1099. While the flare frequency in HR 1099 is lower than that of the highly active ultra-fast rotator AB Dor \citep{2024ApJ...966...86S}, it significantly exceeds the occurrence rate of X-class flares observed on the Sun. However, due to the limited dataset and small sample size for HR 1099, we were unable to plot the flare frequency distribution function as shown in Figure 10 of \cite{2023RAA....23h5017A}. They found a power-law index ($\alpha$) of 2.0$\pm$0.2 for K-type dwarfs, following the relation dN/dE $\propto$ $E^{-\alpha}$. Future research will expand this study to include more K-type subgiants, enhancing statistical significance and deepening our understanding.

The quiescent state of HR 1099 is characterized by a four-temperature plasma, in which the coolest three temperatures represent the quiescent state of HR 1099  with values of 3.02, 6.96, and 12.53 MK and an average temperature of 9.4 MK. The presence of a fourth temperature during the pre-and post-flare state could be due to the presence of flaring components as both flares F1 and F2 are observed one after another. From high-resolution X-ray spectra, the three temperature quiescent state was modeled using {\sc vapec} with temperatures 4.5, 9.6, 20.2 MK with a weighted average temperature of 15.1 MK and average EM of 2.6 $\times$ $10^{53} cm^{-3}$; the average quiescent temperature and EM is found to be similar to the previous results with 4-T quiescent as reported by \cite{2001A&A...365L.318A, 2003A&A...398.1137A} with values nearly 17 MK and 2-T quiescent by \cite{2004ApJS..153..317O} with values 18.6 MK and 4 $\times$ $10^{53} cm^{-3}$. However, \cite{2024MNRAS.528.4591B} reported the 4-T quiescent with lower plasma parameters like the average temperature of 6 MK whereas average EM of 1.3 $\times$ $10^{52} cm^{-3}$.


The TRS of the flaring spectra provides detailed insights into the physical processes occurring within the flaring region and helps to understand the flare evolution. 
According to the general flare evolution scenario, a heat pulse causes a rapid increase in the temperature throughout the loop. This heat is conducted from the hot regions to the cooler chromosphere, causing the chromospheric plasma to expand upward and evaporate explosively. Once the evaporation rate slows down, it balances with cooling, and the plasma cools rapidly through conduction while its density increases. Eventually, radiative cooling becomes more effective than conduction, causing the density to peak and the loop to start to deplete. Radiative cooling then becomes the dominant cooling mechanism \citep[][]{1991A&A...241..197S,2007A&A...471..271R,2014LRSP...11....4R}. The observed increment in coronal loop temperature, plasma density, and abundance resembles the plasma heating and chromospheric evaporation scenario in the observed flaring loops. Meanwhile, during the decay phase of the flare, the decrease in these parameters indicates the cooling mechanism with sustained heating. 
However, in the present case, due to the limited time resolution, we could observe that the temperature peaked either during the rising phase or simultaneously peaked with the luminosity and emission measure or at the peak phase of the flares.  However, the abundances peaked after the emission measure peak, which is evidence of chromospheric evaporation. 

\begin{figure}
\centering

\includegraphics[width=0.93\columnwidth,trim={0.5cm 0.0cm 1.0cm 2.0cm}, clip]{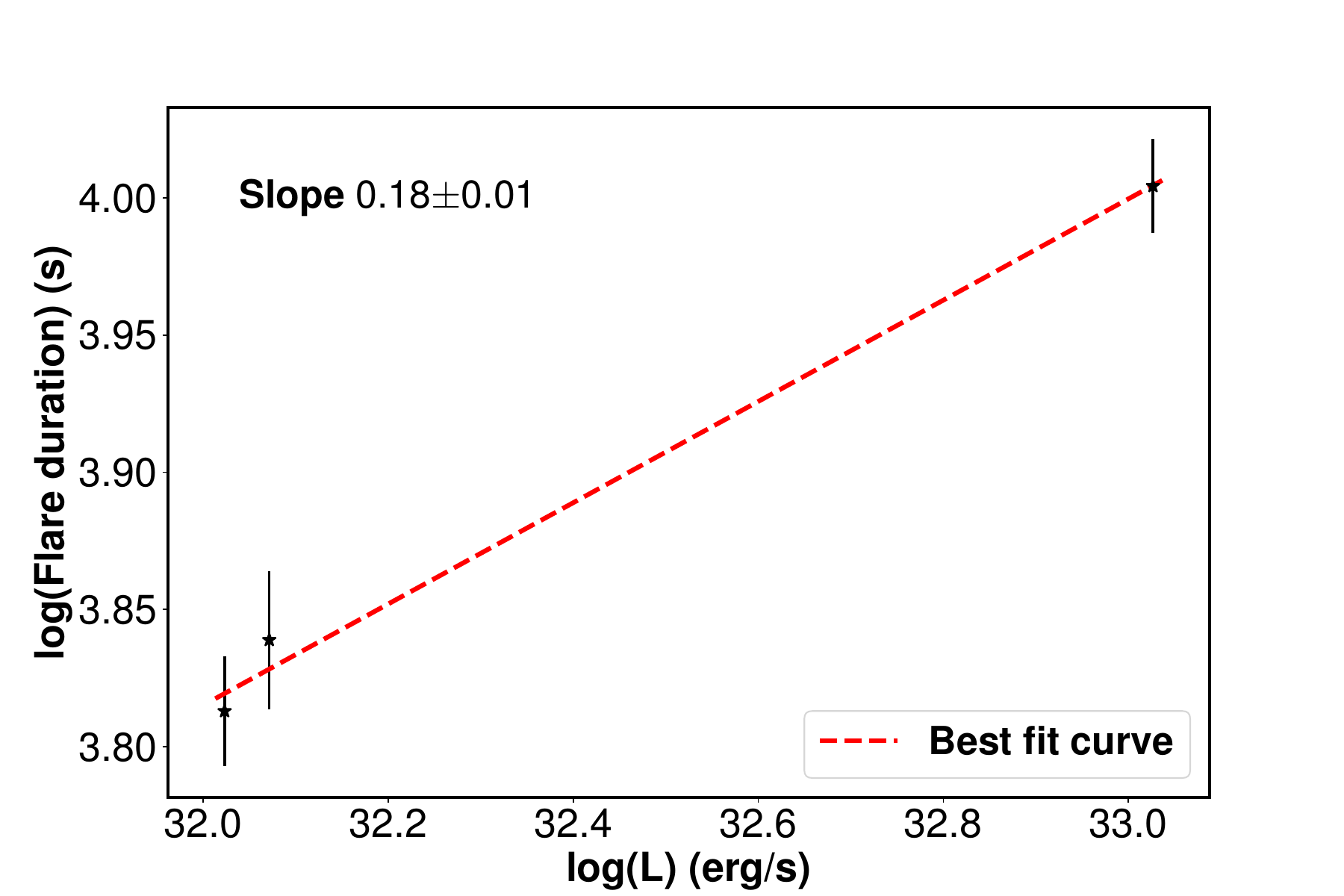}
\caption{Log-log plot of flare duration versus luminosity with linear curve fit.}
\label{fig:L_slope}
\end{figure}

The TRS results show averaged peak temperatures of 39.44, 35.96, and 32.48 MK, emission measures of 7, 7, and $81 \times 10^{53}$ cm$^{-3}$ for flares F1, F2, and F3, respectively. 
Further, we found the global abundances of 0.250, 0.299, and 0.362 $Z_\odot$, peak X-ray luminosities (\lxf) of 1.62, 1.68, and $19.40 \times 10^{31}$ erg $s^{-1}$, and total flare energy of $6.8 \times 10^{35}$, $8.1 \times 10^{35}$, and $1.07 \times 10^{37}$ erg. These results show similarity with the past results for this active star observed with various instruments like EUVE, MAXI, Chandra, Swift, etc. \citep[][]{2006ApJS..164..173M,2007A&A...464..309N,2016PASJ...68...90T,2024AAS...24344706K}.
\cite{2002A&A...382.1070V}, \cite{2008ApJ...677.1385C}, and \cite{2016PASJ...68...90T} established a relationship between flare duration and total flare luminosity, expressed as $\tau$ $\propto$ $L_{XF}^{\alpha}$, where $\alpha$ represents the slope of the log-log plot of flare duration ($\tau$) versus luminosity (\lxf). Based on data from the Sun and nearby stars, they reported slopes of approximately 0.33, 0.2, and 0.2$\pm$0.03 for energy ranges of 3.1-24.8 keV, 6-12 keV, and 0.1-100 keV, respectively. Additionally, \cite{2015EP&S...67...59M} and \cite{2023RAA....23j5010A} reported a similar relationship using optical data from the KEPLER mission. Using our limited data sample, we obtained a slope of 0.18$\pm$0.01 in the energy range of 0.3-10.0 keV, as shown in Figure \ref{fig:L_slope}.

\begin{table*}
    \caption{Abundances of the elements with respect to Fe in the corona of HR 1099  along with the other observations taken from the literature. All the values are expressed relative to the solar photospheric abundances \citep[][]{1989GeCoA..53..197A}.}     \label{tab:abund_all}
    \begin{tabular}{l c c c c c c c c}
         \hline\hline
         Elements($\downarrow$) &  P4   & F3    &   $Z^{a}$ & $Z^{b}$     & $Z^{c}$                      & $Z^{d}$       & $Z^{e}$       & $Z^{f}$  \\
                                &       &       &           &             & ~~~~~~~~Flare /   Quiescent       &\\
         \hline
         $[C/Fe]$     	& 4.4$\pm$0.5 & 4.6$\pm$0.7 &    ...	        & ...           & 4.0$\pm$0.3  /  3.1$\pm$0.3	& ...  	        & 2.75$\pm$0.6	& ... \\
         $[N/Fe]$     	& 2.5$\pm$0.4 & 4.0$\pm$0.6 &    ...	        & ...           & 3.3$\pm$0.2  / 2.6$\pm$0.3	& 4.37$\pm$0.3	& 4.1$\pm$0.8	& ... \\
         $[O/Fe]$     	& 2.7$\pm$0.2 & 2.8$\pm$0.4 &    ...    	    & 3$\pm$0.1	    & 1.81$\pm$0.05	/ 1.5$\pm$0.1	& 2.63$\pm$0.06 & 2.75$\pm$0.5	& 1.75$\pm$0.8 \\
         $[Ne/Fe]$    	& 7.1$\pm$0.4 & 7.2$\pm$0.9 &    8.2$\pm$0.5	& 9.77$\pm$0.1	&  3.4$\pm$0.2	/ 2.7$\pm$0.3	& 7.59$\pm$0.2	& 6.6$\pm$1.1	& 8.67$\pm$2.5 \\
         $[Mg/Fe]$    	& 1.9$\pm$0.2 & 2.0$\pm$0.3 &    2.0$\pm$0.1	& 2.5$\pm$0.1	& 2.2$\pm$0.2	/ 3.0$\pm$0.5	& 1.55$\pm$0.04	& 0.9$\pm$0.3	& 2.42$\pm$0.9 \\
         $[Si/Fe]$    	& 2.5$\pm$0.3 & 2.8$\pm$0.4 &    1.8$\pm$0.1	& 1.9$\pm$0.1	& 2.0$\pm$0.1	/ 2.2$\pm$0.4	& 1.51$\pm$0.04	& 1.1$\pm$0.3	& 2.17$\pm$0.7 \\
         $[S/Fe]$     	& 1.9$\pm$0.5 & 2.3$\pm$0.5 &    1.6$\pm$0.2	& 1.48$\pm$0.2	& 1.6$\pm$0.1   / <1.2           & 1.51$\pm$0.07	& 0.5$\pm$0.4	& 2.58$\pm$0.9 \\
         $[Ar/Fe]$    	& 4.3$\pm$0.9 & 6$\pm$1     &    ...         	& ...           & ...           / ...           & 4.7$\pm$0.4	& 2.5$\pm$1.3	& 4.0$\pm$1.9 \\
         \hline
         \end{tabular}
         ~~\\
         $^a$\cite{2024MNRAS.528.4591B}, $^b$\cite{2001ApJ...548L..81D}, $^c$\cite{2008A&A...482..639N}, $^d$ \cite{2013ApJ...768..135H}, $^e$\cite{2003A&A...398.1137A}, $^f$\cite{2004ApJS..153..317O}\\
\end{table*}

\begin{figure}[b]
\centering

\includegraphics[width=0.99\columnwidth,trim={0.1cm 0.0cm 0.2cm 0.0cm}, clip]{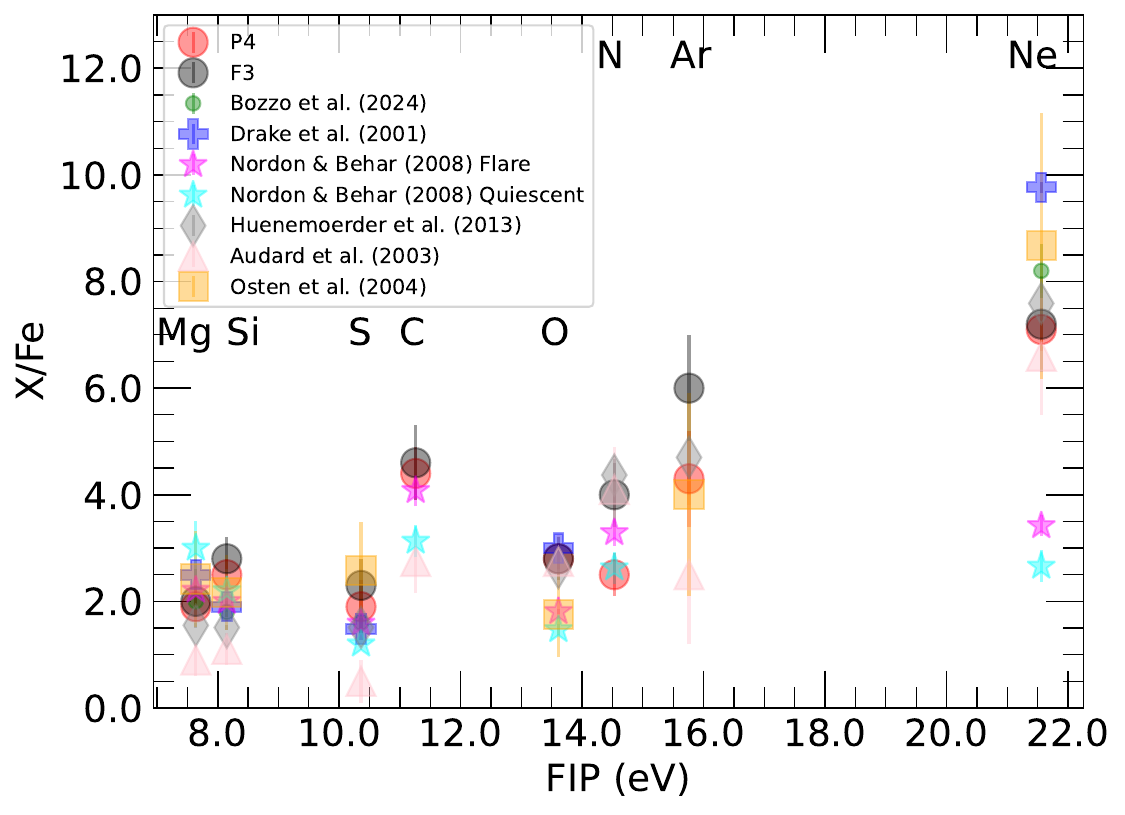}

\caption{All the elemental abundances relative to Fe are plotted against the first ionization potential (FIP). The values are provided in Table \ref{tab:abund_all}.}
\label{fig:fip_all}
\end{figure}
The elemental abundances provide detailed information about the physical nature of the plasma present in the coronae of the stars. The individual abundances during both the quiescent and flaring states in the coronal loops of HR 1099 exhibit the i-FIP effect, a phenomenon also observed in other magnetically active stars \citep[][]{2001A&A...365L.324B,2022A&A...659A...3S}.  
The FIP and i-FIP effects in the active stars have been explained in detail by \cite{2019ApJ...879..124L}. The ponderomotive force associated with the Alfvén waves and fast mode waves separates ions from neutrals in the chromospheric regions, which can explain the physical mechanism responsible for such effects \citep[][and references therein]{2021ApJ...909...17L,2021ApJ...911...86T}. The resonant Alfvén waves propagating along coronal loops can result in the FIP effect, while the i-FIP effect can be reasonably explained by upward-propagating p-modes. The magneto-acoustic waves or p-modes convert to fast modes and propagate into the low plasma beta ($\beta$ < 1) regions \citep[][]{2015LRSP...12....2L,2021ApJ...909...17L,2019ApJ...875...35B}. Such plasma processes can also be attributed as a physical cause of the FIP and i-FIP effects observed presently in the flaring epoch of the localized corona of HR 1099.

The i-FIP effect in HR 1099 has also been previously reported \citep[][]{2001ApJ...548L..81D,2001A&A...365L.324B,2008A&A...482..639N,2022A&A...659A...3S}. In this study, we provide a comparative analysis of the coronal elemental abundances and the i-FIP effect in the coronal loop plasma of HR 1099 with the other estimations existing in the previous literature. The coronal abundances (Z/Z$\odot$) relative to Fe are presented in Table \ref{tab:abund_all} and Figure \ref{fig:fip_all} for P4 and flare F3, alongside previously reported values obtained from various instruments. In this study, we assumed the photospheric values of HR 1099 to be analogous to those of the solar photosphere. Here, we assumed the photospheric values of HR 1099 to be analogous to those of the solar photosphere. The derived values of abundance are found to be consistent with the past literature values.
However, the abundances of Fe and Ne as derived by \cite{2007A&A...464..309N} exhibit lower values compared to those derived in other studies. 
The Fe and O abundances align closely with earlier reports, with minimal uncertainty, making the Fe/O ratio a useful indicator of the FIP bias in coronal abundances \citep[][]{2010ApJ...717.1279W,2012ApJ...753...76W}. A lower Fe/O ratio signifies a stronger i-FIP effect in the star's corona.
The Fe/O abundance ratio during the quiescent state P4 and flaring state F3 of HR 1099 were found to be 0.36$\pm$0.02 and 0.35$\pm$0.05, respectively. These values are consistent with previously reported values between 0.3 and 0.7 for HR 1099, as shown in Table \ref{tab:abund_all}. In other active RS CVn binaries, the Fe/O ratio varies based on the star’s activity level and spectral type \citep[UX Ari, $\lambda$ And, VY Ari, Capella etc;][]{2003A&A...398.1137A,2007A&A...464..309N,2022A&A...659A...3S}.

We derived the semi-loop lengths of the flares using the hydrodynamic loop model, finding them to range between 5.9 and 8.9 $\times 10^{10}$ cm. In comparison, typical loop lengths for solar flares are of the order of 10$^9$ - 10$^{10}$ cm. The loop heights (h = 2L/$\pi$) corresponding to these semi-loop lengths constitute a small fraction of the radius of HR 1099. The h/$R_{*}$ ratios for these flares were found to be 0.14 for both F1 and F2 and 0.21 for the flare F3. \cite{2006ApJS..164..173M} reported the value of L and the ratio h/$R_{*}$ in the range of 4.8-12.7 $\times 10^{10}$ cm and 0.1 to 0.3 for HR 1099. 

The total X-ray energy released during the flares was estimated to be between 6.8 $\times$ 10$^{35}$ and 10.7 $\times$ 10$^{36}$ erg, which is substantially higher than the total energy of the most powerful solar flares 
\citep[][]{2012ApJ...759...71E,2020ApJ...891..138Z}. The calculated loop volume V, electron density $n_{e}$, and plasma pressure P were found to be in the range of 1.3 - 4.4 $\times$ 10$^{31} cm^3$, 2.5 - 4.8 $\times$ 10$^{11} cm^{-3}$, and 5 - 8 $\times$ 10$^{3}$  dyne $cm^{-2}$. Additionally, the total magnetic field strength $B_{Total}$ of the loops was found to be between 500 to 600 G, similar to typical magnetic field observations for HR 1099 \citep[][]{2006ApJS..164..173M, 2012MNRAS.419.1219P} and the minimum magnetic field $B_{min}$ in the range of 350 - 448 G \citep[110-320 G;][]{2006ApJS..164..173M}. 

The mass of ejected CME associated with the flare can be estimated with the relation $M_{CME} (g) = 10^{-1.5\pm0.5} E_{G}^{0.59\pm0.02}$, where $E _{G}$ is the energy released during the flare in GOES (1 -- 8 \AA) energy band \citep{2012ApJ...760....9A,2013ApJ...764..170D}, which can be converted from XMM-Newton flux from 0.3-10.0 keV band to GOES flux using WEBPIMMS. The $M_{CME}$ for HR 1099 was found to be in the range of approximately 3.0 - 14.5 $\times$ $10^{19}$ g, which was found to be maximum for flare F3 as mentioned in Table \ref{tab:final_para}. These values are approximately 100 to 1000 times higher than the most massive solar CME reported \citep{2009IAUS..257..233Y}. However, they exceed the mass of CMEs observed on other active stars such as AB Dor, EQ Peg, EK Dra, etc., which have typical CME masses in the range of $10^{18-19}$ g \citep{2024MNRAS.527.1705D, 2022MNRAS.509.3247K, 2022NatAs...6..241N}. Based on the calculated parameters for HR 1099 using X-ray data, such as flare temperature, emission measure, loop length, magnetic field, etc., it is evident that the study of flare parameters plays a crucial role in understanding the coronal dynamics of the magnetically active stars. The more resolved data and multi-wavelength perspective can provide a comprehensive understanding of the geometry of the corona and the physical processes happening during these flares in detail.

\section*{Acknowledgements}
This work presented here is based on the observations of the XMM–Newton satellite, an ESA science mission with instruments and contributions directly funded by ESA Member States and NASA. SD acknowledges the CSIR funding agency for providing the research grant. AKS acknowledges the support of the ISRO project in facilitating this scientific research. We thank the reviewer for his/her valuable comments and suggestions. 
\section*{Data Availability}
The data utilized in this paper is accessible through the XMM-Newton archive.

\bibliography{HR1099AJ}{}
\bibliographystyle{aasjournal}

\end{document}